\documentclass[12pt]{article}
\usepackage{color}
\usepackage{latexsym}
\usepackage{epsfig,amssymb,euscript, mathrsfs}
\usepackage{amsmath}
\usepackage{graphicx}
\newsavebox{\ns}
\newsavebox{\dbrane}
\newsavebox{\dbshort}

\def\be{\begin{equation}}
\def\ee{\end{equation}}
\def\bea{\begin{eqnarray}}
\def\eea{\end{eqnarray}}

\newcommand{\nn}{\nonumber}

\newcommand\R{\mathbb{R}}
\newcommand\Z{\mathbb{Z}}

\newcommand\C{\mathbb{C}}

\newcommand\diff{\mathrm{d}}
\newcommand{\de}{\partial}

\newcommand{\dd}{\mathrm{d}}

\newcommand{\ii}{\mathrm{i}}

\newcommand{\ex}{\mathrm{e}}
\newcommand{\vol}{\mathrm{vol}}
\newcommand{\Vol}{\mathrm{Vol}}
\newcommand{\Tr}{\mathrm{Tr}}

\newcommand{\Imag}{\mathrm{Im}\, }
\newcommand{\Real}{\mathrm{Re}\, }

\newcommand\e{\epsilon}

\newcommand{\xmax}{{x_\mathrm{max}}}
\newcommand{\xmin}{{x_\mathrm{min}}}

\newcommand{\eps}{\varepsilon}
\newcommand{\et}{\varepsilon}
\newcommand{\FF}{\mathscr{F}}

\newlength{\sswidth}

\newcommand{\gauge}{g}
\newcommand{\ssigma}{\eta}
\newcommand{\cmax}{c_{\mathrm{max}}}
\newcommand{\cmin}{c_{\mathrm{min}}}
\newcommand{\n}{P}

\numberwithin{equation}{section}       

\begin{document}

\begin{titlepage}

\begin{center}

\today

\vskip 2.3 cm 

{\Large \bf  Wilson loops and the geometry of}
\vskip .5cm 
{\Large \bf matrix models in AdS$_4$/CFT$_3$} 

\vskip 2 cm

{Daniel Farquet and James Sparks\\}

\vskip 1cm

\textit{Mathematical Institute, University of Oxford,\\
24-29 St Giles', Oxford OX1 3LB, United Kingdom\\}

\end{center}

\vskip 2 cm

\begin{abstract}
\noindent We study a general class of supersymmetric AdS$_4\times Y_7$ solutions 
of M-theory that have large $N$ dual descriptions as $\mathcal{N}=2$ Chern-Simons-matter theories on $S^3$. 
The Hamiltonian function $h_M$ for the M-theory circle, with respect to a certain contact structure on $Y_7$, 
plays an important role in the duality. We show that an M2-brane wrapping the M-theory circle, 
giving a fundamental string in AdS$_4$, is supersymmetric precisely at the critical points 
of $h_M$, and moreover the value of this function at the critical point determines the 
M2-brane action. Such a configuration determines the holographic dual of a 
BPS Wilson loop for a Hopf circle in $S^3$, and leads to an effective method 
for computing the Wilson loop on both sides of the correspondence in large classes of examples. 
We find agreement in all cases, including for several infinite families, 
and moreover we find that the image $h_M(Y_7)$ determines the range of 
support of the eigenvalues in the dual large $N$ matrix model, with the 
critical points of $h_M$ mapping to points where the derivative of the eigenvalue density is discontinuous.
 
\end{abstract}

\end{titlepage}

\pagestyle{plain}
\setcounter{page}{1}
\newcounter{bean}
\baselineskip18pt

\tableofcontents

\section{Introduction and summary}

Over the last few years our understanding of the AdS$_4$/CFT$_3$ correspondence, particularly in 
 M-theory, has improved considerably. Broadly speaking, this has involved developments 
on two fronts. Firstly,  we now have large classes of very explicit examples of dual pairs; 
that is, gravity backgrounds for which we have some precise description of the dual  superconformal field theories. Secondly, 
there are new \emph{quantitative} tests of these conjectured dualities, based on supersymmetric localization in the
field theories. The aim of this article is to extend this quantitative analysis further, by examining the computation 
of certain BPS Wilson loops on both sides of the correspondence. In the process we will also understand 
how other structures are related via the duality.

 Starting with 
the seminal work of \cite{ABJM} we now have large classes of supersymmetric
 AdS$_4\times Y_7$ 
gravity backgrounds of M-theory that are associated with particular $(2+1)$--dimensional  supersymmetric  gauge theories, typically Chern-Simons theories coupled to matter, that are believed
to have a dual superconformal fixed point. The construction of the
UV gauge theory usually relies on a dual description in terms of type IIA string theory, 
which in turn involves a choice of M-theory circle $U(1)_M$ acting on $Y_7$; different choices 
can lead to different UV gauge theories that flow to the same 
IR superconformal fixed point. In \cite{ABJM} the highly supersymmetric
case where $Y_7=S^7/\Z_k$, equipped with its round Einstein metric and with $N$ units of flux
through this internal space, 
was related to a large $N$ dual description as an $\mathcal{N}=6$ superconformal $U(N)\times U(N)$ Chern-Simons-matter 
theory (the ABJM theory), with $k\in\Z$ being the Chern-Simons coupling. Here $\Z_k\subset U(1)_M$, 
with the M-theory circle action by $U(1)_M$ being the Hopf action 
on $S^7$, so that $S^7/U(1)_M=\mathbb{CP}^3$.
There are now many families of examples of a similar type \cite{Benna:2008zy}--\cite{Closset:2012ep}, generally 
with $\mathcal{N}\geq 2$ supersymmetry, in which $Y_7$ is a Sasaki-Einstein seven-manifold and 
the dual description typically involves supersymmetric Chern-Simons-matter
theories whose gauge groups are products of unitary groups, 
and with matter in various representations (bifundamental, fundamental, adjoint). There are 
also examples in which AdS$_4\times Y_7$ is a warped product, 
with non-trivial four-form flux on non-Einstein $Y_7$ (obtained 
thus far either by marginal \cite{Lunin:2005jy} or relevant \cite{Corrado:2001nv}--\cite{Gabella:2012rc} deformations  of Einstein examples). 

Quantitative tests of these conjectured dualities arise by putting 
the (Euclidean) field theories on a compact three-manifold. 
The simplest case, in which this three-manifold 
is taken to be $S^3$ equipped with its round metric, 
was studied in \cite{Kapustin:2009kz, Jafferis:2010un, Hama:2010av}. This 
can be done for a completely general $\mathcal{N}=2$ supersymmetric 
gauge theory, in such a way to preserve supersymmetry. Moreover, 
using a standard argument \cite{Pestun:2007rz} one can show that 
the path integral, with any BPS operator inserted, reduces exactly
to a finite-dimensional matrix integral. This implies that the 
VEVs of BPS operators may be computed exactly using 
a matrix model description, with the large $N$ limit of 
this then expected to reproduce certain supergravity results. 
In practice this has been used to compute the free energy  $F$
(minus the logarithm of the partition function) on both sides 
of the correspondence \cite{Herzog:2010hf}--\cite{Amariti:2012tj}, where on the supergravity side 
this is proportional to $N^{3/2}$ with a coefficient 
depending only on the volume of $Y_7$.\footnote{For a general  AdS$_4\times Y_7$ solution this is the \emph{contact volume} 
of $Y_7$, rather than the Riemannian volume, as we shall review in section \ref{sec:M2}.}

It is natural to try to extend these results further, by inserting non-trivial BPS operators into the 
path integral, computing the corresponding large $N$ behaviour in the matrix model, 
and comparing to an appropriate dual semi-classical supergravity computation. 
In the original papers on the ABJM theory \cite{Kapustin:2009kz, Herzog:2010hf, Drukker:2009hy, Marino:2009jd, Drukker:2010nc, Suyama:2009pd} 
the supersymmetric Wilson loop for the gauge field around a Hopf circle $S^1\subset S^3$ was studied. This 
is 1/2 BPS, and is readily computed in the large $N$ matrix model \cite{Kapustin:2009kz, Herzog:2010hf}. 
Generally speaking, one expects such a Wilson loop to be dual to a fundamental string 
when viewed from a type IIA perspective  \cite{Maldacena:1998im}, with the Euclidean string worldsheet 
having boundary on the Hopf $S^1$ at conformal infinity. Semi-classically, more precisely 
this will be a supersymmetric minimal surface $\Sigma_2$ in Euclidean AdS$_4$, with the VEV calculated 
via the regularized area of the string worldsheet.
Such a string must then be pointlike in the internal space, and 
for the ABJM theory this is $\mathbb{CP}^3=S^7/U(1)_M$. 
Equivalently,  this IIA string lifts to an M2-brane wrapping the M-theory circle.
Notice that since $\mathbb{CP}^3$ is a homogeneous space all positions for the IIA string 
are equivalent.
The two computations (large $N$ matrix model and area) of course agree.\footnote{Similar 
Wilson loops have recently been considered in five-dimensional superconformal field
theories on $S^5$ \cite{Assel:2012nf}, which may also be computed using
localization techniques. The gravity duals are described by warped AdS$_6 \times
S^4/\Z_n$ solutions of massive IIA supergravity, and thus the geometry of the
internal spaces here is fixed and in fact unique \cite{Passias:2012vp}.}

\begin{figure}[ht!]
\centering
\includegraphics[width=0.56\textwidth]{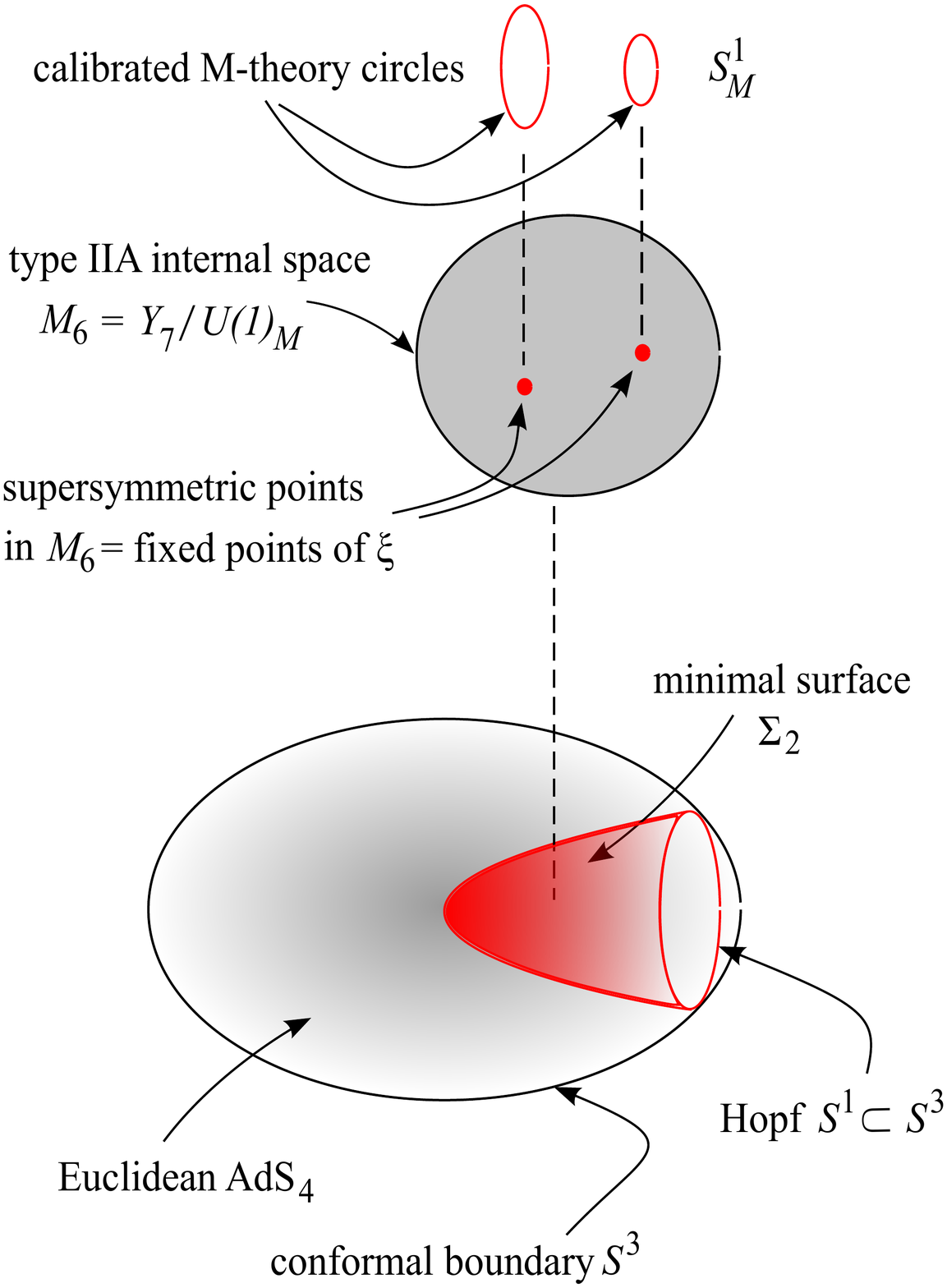}
\caption{A depiction of the total spacetime AdS$_4\times Y_7$, with a choice 
of M-theory circle $U(1)_M$, together with the supersymmetric M2-branes of interest which are shown in red. These M2-branes are
pointlike in the type IIA internal space $M_6=Y_7/U(1)_M$, 
wrapping copies of the M-theory circle over these points, and are calibrated by the contact 
form $\ssigma$. The supersymmetric points in $M_6$ are precisely the points where the 
projection of the R-symmetry/Reeb vector field $\xi$ is zero (giving fixed points on $M_6$), and in general the calibrated 
circles over such points have different lengths. 
The remaining worldvolume of the M2-brane wraps a minimal supersymmetric 
surface $\Sigma_2$ in Euclidean AdS$_4$. The latter may be viewed 
as a hyperbolic 4-ball, with conformal boundary $S^3$, 
and $\Sigma_2$ then has the topology of a 2-ball, with 
boundary a Hopf $S^1\subset S^3$.}\label{fig1}
\end{figure}

This Wilson loop is 1/2 BPS in a general $\mathcal{N}=2$ supersymmetric 
gauge theory on $S^3$, as we review in section \ref{sec:Wilson},
and can be computed using the large $N$ matrix model description. 
The supergravity dual computation will naturally involve an M2-brane wrapping 
the M-theory circle, leading to the same fundamental string configuration in 
Euclidean AdS$_4$ (see Figure \ref{fig1}). The only issue is which copy of the M-theory 
circle is relevant? When the internal space is $Y_7=S^7/\Z_k$ all 
choices are equivalent by symmetry, but on a general Sasaki-Einstein manifold $Y_7$, 
or a more general non-Einstein $Y_7$ with flux, this is clearly not the case. 
Equivalently we may ask which IIA fundamental strings in AdS$_4\times M_6$, that are pointlike in $M_6=Y_7/U(1)_M$,
preserve any supersymmetry. 

\subsubsection*{Summary of results}

Our starting point is to consider BPS M2-branes in general $\mathcal{N}=2$ supersymmetric 
AdS$_4\times Y_7$ solutions of eleven-dimensional supergravity. These backgrounds 
were studied in detail in \cite{Gabella:2011sg, Gabella:2012rc}, where it was shown 
that provided the quantized M2-brane charge $N$ of the background (measured 
by a certain flux integral) is non-zero, then there is always a canonical \emph{contact one-form} 
$\ssigma$ defined on $Y_7$. Concretely, $\ssigma$ is constructed as a bilinear in the Killing spinors 
on $Y_7$, and it was shown in the latter reference that this contact structure 
entirely captures both the gravitational free energy of the background, and 
also the scaling dimensions of BPS operators arising from supersymmetric M5-branes 
wrapped on five-manifolds $\Sigma_5\subset Y_7$. 

In this paper we will show that 
the same contact form $\ssigma$ captures the Wilson loop VEV $\langle W \rangle$ of interest, computed semi-classically
from the action of a BPS M2-brane.
 Concretely, we derive the general formula
\bea\label{Wgravity}
\log\ \langle W \rangle_{\mathrm{gravity}} &=& \frac{(2\pi)^2\int_{S^1_M}\ssigma}{\sqrt{96\, \Vol_\ssigma(Y_7)}}N^{1/2}~,
\eea
where we have defined the \emph{contact volume} of $Y_7$ as
\bea\label{contactvol}
\Vol_\ssigma(Y_7) &\equiv & \frac{1}{48}\int_{Y_7}\ssigma\wedge (\diff\ssigma)^3~.
\eea
In particular, a supersymmetric M2-brane is \emph{calibrated} with respect to $\ssigma$, which is 
why the integral of $\ssigma$ along the M-theory circle $S^1_M$ appears in the formula (\ref{Wgravity}). 
A contact form $\ssigma$ always has an associated unique \emph{Reeb vector field} $\xi$, defined
via the equations $\xi\lrcorner\ssigma=1$, $\xi\lrcorner\diff\ssigma=0$, and in 
\cite{Gabella:2011sg, Gabella:2012rc} it was shown that $\xi$ is also the 
\emph{R-symmetry} Killing vector field, that is expected since 
an $\mathcal{N}=2$ superconformal theory in three dimensions 
has a $\mathtt{u}(1)_R$ symmetry in the superconformal algebra. 
We will show that an M2-brane wrapping a copy of the M-theory circle $S^1_M$ is supersymmetric
precisely when the generating vector field $\zeta_M$ of $U(1)_M$ is proportional 
to $\xi$. Geometrically, this means that the corresponding fundamental string at a point $p\in M_6$ 
is supersymmetric precisely when $p$ is a fixed point of $\xi$, considered as a vector field on $M_6$ 
(on $Y_7$, on the other hand, $\xi$ is always nowhere zero).

There is another way to describe which wrapped M2-branes are supersymmetric which involves 
the \emph{Hamiltonian function} for the M-theory circle, defined as
\bea
h_M &\equiv & \ssigma(\zeta_M)~.
\eea
This is a real function $h_M:Y_7\rightarrow \R$, invariant under $\zeta_M$, and we show that 
the supersymmetric M-theory circles $S^1_M\subset Y_7$ lie precisely on the critical set 
$\diff h_M=0$. The action of a supersymmetric M2-brane corresponding to a point $p\in M_6$ may then also be written as
\bea\label{M2action}
-S_{\mathrm{M2}} &=& \frac{(2\pi)^3h_M(\hat{p})}{\sqrt{96\, \Vol_\ssigma(Y_7)}}N^{1/2}~,
\eea
where $\hat{p}\in Y_7$ is any point that projects to $p\in M_6=Y_7/U(1)_M$. 
Since (\ref{M2action}) depends only on $\ssigma$ we may compute this expression in examples
using the same methods employed in \cite{Gabella:2011sg, Gabella:2012rc}, \cite{Gabella:2010cy}--\cite{Martelli:2006yb}. For example, 
for toric solutions (\ref{M2action}) may be computed entirely using toric geometry methods. 
In general there are multiple supersymmetric $S^1_M$  circles, 
which can have different lengths with respect to $\ssigma$ and thus leading to different actions (\ref{M2action}).
 In the semi-classical computation
one should \emph{sum} over all such configurations, which in the large $N$ limit then implies that in (\ref{Wgravity}) 
it is the \emph{longest} $S^1_M$  that gives the leading contribution to the Wilson loop.

In the families of examples that we shall study, the dual field theory computation 
of the Wilson loop VEV reduces to a computation in a  large $N$ matrix model. 
As we shall review in section \ref{sec:Wilson}, in this matrix model the eigenvalues at large $N$
take the general form $\lambda^I = x N^{1/2}  + \ii y^I(x)$, where the index $I$ runs 
over the number of factors of $U(N)$ in the gauge group $G=\prod_I U(N)$,
and are described by an eigenvalue density function $\rho(x)$ which is supported 
on some interval $[\xmin,\xmax]\subset \R$. To leading order at large $N$ it is straightforward to compute
\bea\label{Wxmax}
\log \ \langle W \rangle_{\mathrm{QFT}} &=& \xmax\, N^{1/2}~,
\eea
which should be compared to the dual supergravity result (\ref{Wgravity}). 

Remarkably,  in all examples that we study we find that the interval $[\xmin,\xmax]$ in the matrix model 
coincides, in a precise way, with the image of the Hamiltonian function $h_M(Y_7)$. Since $Y_7$ is compact 
and connected, the latter image is also necessarily a closed interval, and more precisely we find 
$h_M(Y_7)=[\cmin,\cmax]$, where the field theory quantity $x$  is proportional to  the geometrical quantity $c$: 
\be\label{xc}
x \ = \ \frac{(2\pi)^3}{\sqrt{96\, \Vol_\ssigma(Y_7)}}\, c~.
\ee
The Hamiltonian $h_M$ is a Morse-Bott function on the symplectic cone over $Y_7$, and on general grounds we know that 
the image interval  $[\cmin,\cmax]$ is divided into $\n$ subintervals 
$\cmin = c_1< c_2 < \cdots < c_{\n+1}=\cmax$, where the critical set 
maps as
 $h_M\left(\{\diff h_M=0\}\right) =\{c_i\mid i=1,\ldots,\n+1\}$.
For all $c\in (c_i,c_{i+1})$ 
the level surfaces $h_M^{-1}(c)\subset Y_7$ are diffeomorphic to a fixed six-manifold, with 
the topology changing precisely as one passes a critical point $c_i$. 
Even more remarkable is that we find that the corresponding 
points $x_i$, related to $c_i$ via (\ref{xc}),  
are precisely the points where $\rho'(x)$ has a jump discontinuity in the matrix model. 
These points are then also related to the fixed points of the Reeb vector $\xi$ on $M_6$.

The outline of the rest of this article is as follows. In section \ref{sec:Wilson} we review 
the definition of the BPS Wilson loop in $\mathcal{N}=2$ Chern-Simons-matter theories, 
and  how it may be computed in the large $N$ matrix model. Section \ref{sec:M2} 
analyses supersymmetric M2-branes in a general class of 
AdS$_4\times Y_7$ backgrounds in M-theory, and we derive the general formula 
for the action (\ref{M2action}),  leading to the holographic Wilson loop result (\ref{Wgravity}). 
In section \ref{sec:examples} we compute the Wilson loop, on both sides 
of the correspondence, in a variety of examples, including for
several infinite families of Sasaki-Einstein $Y_7$, and for models 
with non-Einstein $Y_7$. Section \ref{sec:discussion} concludes 
with a brief discussion.

\section{Wilson loops in $\mathcal{N}=2$ gauge theories on $S^3$}\label{sec:Wilson}

The dual superconformal field theories of interest  have UV descriptions as $\mathcal{N}=2$ Chern-Simons gauge theories with matter on $S^3$. 
We begin in this section by defining the BPS Wilson loop in such a theory,  summarize how it localizes in the matrix model, and explain how it can be efficiently calculated. This section is mainly a review of material already in the literature.

\subsection{The Wilson loop}\label{sec:WL}

 In $\mathcal{N}=2$ supersymmetric gauge theories the gauge field $A_\mu$ is part of a vector multiplet that also contains two real scalars $\sigma$ and $D$, that are auxiliary fields, and a two-component spinor $\lambda$, all of which are in the adjoint representation of the gauge group $G$. 
The BPS Wilson loop in a representation $\mathcal{R}$ of  $G$ is given by
\bea\label{defW}
W & =& \frac{1}{\dim \mathcal{R}}\Tr_\mathcal{R}\left[\mathcal{P}\exp \left(\oint_\gamma \dd s(\ii A_\mu\dot x^\mu+\sigma|\dot x|)\right)\right]~,
\eea
where $x^\mu(s)$ parametrizes the worldline $\gamma\subset S^3$ of the Wilson line and the path ordering operator has been denoted by $\mathcal{P}$. 
For a Chern-Simons theory the gauge multiplet has a kinetic term described by the supersymmetric Chern-Simons action
\bea\label{CS}
S_{\mathrm{Chern}-\mathrm{Simons}} \ = \ \frac{k}{4\pi}\int \dd^3x \sqrt{\det g}\ \Tr\left(A\wedge\dd A+\frac{2}{3}A\wedge A\wedge A-\lambda^\dag\lambda+2D\sigma\right)~,
\eea
where here $g$ is the round metric on $S^3$, and $k$ denotes the Chern-Simons coupling. 
When $G$ is a product of unitary groups, $G=\prod_I U(N_I)$, one can in general take different $k_I\in\Z$ for each factor. 
In this case we will denote $k=\mathrm{gcd}\{k_I\}$ \cite{Martelli:2008si}.

There are four Killing spinors on $S^3$, two satisfying each choice of sign in the equation 
$\nabla_\mu \et = \pm \frac{\ii}{2}\tau_\mu\et$, where the gamma matrices $\tau_\mu$ 
in an orthonormal frame generate
the Clifford algebra $\mathrm{Cliff}(3,0)$, and may thus be taken to be the Pauli matrices. 
A natural orthonormal frame $\{e^m\}_{m=1,2,3}$ on $S^3$ is provided by the left (or right) invariant 
one-forms under the isometry group $SU(2)_{\mathrm{left}}\times SU(2)_{\mathrm{right}}$. 
The four Killing spinors on $S^3$ transform in the $(\mathbf{2},\mathbf{1})$, $(\mathbf{1},\mathbf{2})$ 
representations of this group.

The full supersymmetry transformations for 
a vector multiplet and matter multiplet may be found in \cite{Kapustin:2009kz, Jafferis:2010un, Hama:2010av}. 
For our purposes we need note only that localization of the path integral, discussed in the next section, requires 
one to choose a Killing spinor $\et$, which without loss of generality we assume solves 
$\nabla_\mu \et =\frac{\ii}{2}\tau_\mu\et$. This choice of Killing spinor then 
 has the two associated supersymmetry transformations
\bea\label{SUSY}
\nonumber\delta A_\mu&=& -\frac{\ii}{2}\lambda^\dag\tau_\mu\et~,\\
\delta\sigma&=&-\frac{1}{2}\lambda^\dag\et~.\eea
If one varies the Wilson loop \eqref{defW} under the latter supersymmetry transformation one obtains
\bea
\delta W  & \propto & \frac{1}{2}\lambda^\dag(\tau_\mu\dot x^\mu-|\dot x|)\et~.
\eea
The Wilson loop is then invariant under supersymmetry provided
\bea
(\tau_\mu\dot x^\mu-|\dot x|)\et & = & 0~.
\eea
Choosing $s$ to parametrize arclength, so that $|\dot x|=1$ along the loop, we see that $\tau_\mu\dot x^\mu$ must be constant.
 In the left-invariant orthonormal frame $e^m$ one may then align $\dot x^\mu$ along one direction, say $e^3$. The integral 
 curve of this vector field is a Hopf $S^1\subset S^3$ (or equivalently a great circle). The supersymmetry condition then becomes
 \bea\label{Wsusycondition}
 (\tau_{3}-1)\et & =& 0~.
 \eea
This projection condition then fixes one of the two possible choices of $\et$ satisfying 
$\nabla_\mu \et = \frac{\ii}{2}\tau_\mu\et$, implying that  the Wilson loop \eqref{defW} is indeed a $1/2$ BPS operator provided 
one takes $\gamma$ to be a Hopf circle.  We will see later on that the condition \eqref{Wsusycondition}, plus the fact that the supersymmetry generators are Killing spinors, also arises as the 
condition for supersymmetry of a probe M2-brane.

\subsection{Localization in the matrix model}\label{sec:localize}

The VEV of the BPS Wilson loop (\ref{defW}) is, by definition, obtained by inserting $W$ into the path integral 
for the theory on $S^3$. The computation of this is  greatly simplified by the fact that this path integral 
\emph{localizes} onto supersymmetric configurations of fields. We  summarize the main steps and results in this section, 
following in particular \cite{Kapustin:2009kz, Jafferis:2010un, Herzog:2010hf, Martelli:2011qj}, and refer the reader to the original 
papers for further details.

The central idea is that the path integral, with $W$ inserted, is invariant under the supersymmetry variation 
$\delta$ corresponding to the Killing  spinor $\et$ satisfying (\ref{Wsusycondition}). We have written two 
of the supersymmetry variations in (\ref{SUSY}), and the variations of other fields (including fields in the chiral matter multiplets) 
may be found in the above references. Crucially, $\delta^2=0$ is nilpotent. There is then a form of \emph{fixed point theorem} 
that implies that the only net contributions to this path integral come from field configurations 
that are invariant under $\delta$ \cite{Witten:1991zz}. Formally, one can argue this by introducing a collective 
Grassmann coordinate $\vartheta$ along the direction defined by $\delta$ in field space, 
and then appeal to the fact that the Grassmann integral $\int \diff\vartheta=0$. This 
then breaks down precisely at fixed points of $\delta$, where  the coordinate $\vartheta$ is 
not defined. 

Alternatively, and more practically for computation, one may add a conveniently chosen $\delta$-exact positive definite term 
to the action, which a standard  argument shows does not affect the expectation value of any supersymmetric ($\delta$-invariant) 
operator. For  the vector multiplet one can add the term $t\Tr [(\delta\lambda)^\dag\delta\lambda]$ to the action (a similar 
term exists for a matter multiplet), 
without affecting the path integral. Sending $t\rightarrow\infty$
one notes that, due to the form of this term added to the Lagrangian, only configurations with $\delta\lambda=0$ contribute to the path integral in  a saddle point approximation. This saddle point then gives the same value as if the path integral had been calculated with $t=0$, which is the quantity we are interested in. The saddle point approximation requires one to compute a one-loop determinant around the $\delta$-invariant 
field configurations, which in the terminology of fixed point theorems
is the  contribution from the normal bundle to the fixed point set in field space.

For the $\mathcal{N}=2$ supersymmetric Chern-Simons-matter theories of interest, one finds that 
the $\delta$-invariant configurations on $S^3$ are particularly simple: 
\bea\label{localized}
A_\mu &  =& 0~, \quad\text{and}\quad D \ = \ -\sigma \ = \ \text{constant}~,
\eea
with all fields in the matter multiplet set identically to zero.  Here we may diagonalize 
$\sigma$ by a gauge transformation. For a $U(N)$ gauge group we may thus write
$\sigma=\mathrm{diag}(\frac{\lambda_1}{2\pi},\ldots
\frac{\lambda_N}{2\pi})$,  thus parametrizing $2\pi\sigma$ by its eigenvalues $\lambda_i$. The theories of interest will have 
a product gauge group of the form $G=\prod_{I=1}^\gauge U(N)$, 
and for $t=\infty$ the partition function then takes the saddle point form
\bea\label{Z}
Z &=& \frac{1}{ (N!)^\gauge}\int\left(\prod_{I=1}^\gauge \prod_{i=1}^N \frac{\diff\lambda^I_i}{2\pi}\right)\exp\left[\ii \sum_{I=1}^\gauge \frac{k_I}{4\pi}\sum_{i=1}^N (\lambda^I_i)^2\right]\ex^{-F_{\mathrm{one-loop}}}~,
\eea
where the one-loop determinant is given by
\bea\label{oneloop}
\ex^{-F_{\mathrm{one-loop}}} &=& \prod_{I=1}^\gauge \prod_{i\neq j} 2\sinh \frac{\lambda^I_i-\lambda^I_j}{2}\cdot \prod_{\mathrm{matter}\,  \alpha} 
\mathrm{det}_{\mathrm{\mathcal{R_\alpha}}} \exp\left[\ell(1-\Delta_\alpha+\ii \sigma)\right]~.
\eea
Here the first exponential term in (\ref{Z}) is simply the classical Chern-Simons action in (\ref{CS}), evaluated on the 
localized constant field configuration (\ref{localized}). The one-loop determinant factorizes, and 
the first term in (\ref{oneloop}) is the one-loop determinant 
for the vector multiplet. Since we have used gauge-invariance  in (\ref{Z}) to restrict the integral 
to the Cartan subalgebra, we also have a Vandermonde determinant which has been cancelled against a term that appears in the 
one-loop determinant. The second term in (\ref{oneloop}) involves a product over chiral matter multiplets, labelled by $\alpha$. 
We have taken the $\alpha^{\mathrm{th}}$ multiplet to be in representation $\mathcal{R}_\alpha$, and with  R-charge $\Delta_\alpha$. 
The determinant in the representation $\mathcal{R}_\alpha$ 
is understood to be a product over weights $\varrho$ 
in the weight-space decomposition of this representation, and $\sigma$
is then understood
to mean $\varrho(\sigma)$ in (\ref{oneloop}). Finally, 
\bea
\ell(z) &=& -z\log\left(1-\ex^{2\pi \ii z}\right) + \frac{\ii}{2}\left[\pi z^2 + \frac{1}{\pi}\mathrm{Li}_2\left(\ex^{2\pi \ii z}\right)\right]-\frac{\ii\pi}{12}~.
\eea

In this set-up, the VEV of the BPS Wilson loop (\ref{defW}) reduces to
\be\label{W}
\langle W \rangle \ =\ \frac{1}{Z(N!)^g\dim\mathcal{R}} \int\left(\prod_{I=1}^\gauge \prod_{i=1}^N \frac{\diff\lambda^I_i}{2\pi}\right)\ex^{\ii \sum_{I=1}^\gauge \frac{k_I}{4\pi}\sum_{i=1}^N (\lambda^I_i)^2}\, \mathrm{Tr}_{\mathcal{R}} \left(\ex^{2\pi \sigma}\right) \ex^{-F_{\mathrm{one-loop}}}~.
\ee
Notice the integrand is the same as that for the partition function (\ref{Z}), with an additional insertion 
of $\mathrm{Tr}_\mathcal{R}(\ex^{2\pi\sigma})$ arising from the Wilson loop operator. 
The factor of $(N!)^\gauge$, as in (\ref{Z}), arises from dividing by residual Weyl transformations, 
which for $U(N)$ introduces a factor of $1/N!$. Note also that we have normalized the VEV relative 
to the partition function $Z$, so that $\langle 1 \rangle =1$, as is usual in quantum field theory.

Localization has reduced the partition function $Z$ and the Wilson loop VEV 
to finite-dimensional integrals  (\ref{Z}), (\ref{W})  over the eigenvalues $\lambda_i^I$ of $\sigma$,  
but in practice these are difficult to evaluate explicitly due to the complicated one-loop 
effective potential (\ref{oneloop}). For comparison to the dual supergravity 
results we must take the $N\rightarrow\infty$ limit, where the number of eigenvalues, 
and hence integrals, tends to infinity. One can then attempt to compute this 
limit using a saddle point approximation of the integral (this is then our second application of 
the saddle point method). 
With the exception 
of the $\mathcal{N}=6$ supersymmetric ABJM theory, where this matrix model is well-understood \cite{Klemm:2012ii}, 
for general $\mathcal{N}=2$ theories the large $N$ limit of the matrix integrals 
is not understood rigorously. However, in \cite{Herzog:2010hf} a simple 
\emph{ansatz} for the large $N$ limit of the saddle point eigenvalue distribution 
was introduced. This ansatz is based on a partial analytic analysis of the matrix model, 
and also on a numerical approach to computing the saddle point. One seeks saddle points 
with eigenvalues of the form
\bea\label{ansatz}
\lambda^I_i & = & x_i N^{\beta}+\ii y_i^I~,
\eea
with $x_i$ and $y_i^I$ real and assumed to be $\mathcal{O}(1)$ in a large $N$ expansion, and $\beta>0$. In the large $N$ 
limit the real part is assumed to become dense. Ordering the eigenvalues so that the $x_i$ are strictly increasing, 
the real part becomes a continuous variable $x$, with density $\rho(x)$, while $y_i^I$ becomes a continuous function of $x$, $y^I(x)$.

Substituting this ansatz into the partition function expression (\ref{Z}), the sums over 
eigenvalues become Riemann integrals over $x$, and one finds that the \emph{double sums} 
appearing in the one-loop expression (\ref{oneloop}) effectively have a delta function 
contribution which reduces them to single integrals over $x$. (This is often described 
by saying that the \emph{long range forces} in the matrix model cancel.) 
Writing $Z=\ex^{-F}$ one then obtains a functional $F[\rho(x),y^I(x)]$, with 
$x$ supported on some interval $[\xmin,\xmax]$, and to apply the saddle point 
method one then extremizes $F$ with respect to $\rho(x)$, $y^I(x)$, subject
to the constraint that $\rho(x)$ is a density
\bea\label{density}
\int_{\xmin}^{\xmax}\rho(x)\diff x &=& 1~.
\eea
The existence of such a saddle point fixes the exponent $\beta=\frac{1}{2}$ in (\ref{ansatz}).
One then finally also extremizes over the choice of interval, by varying 
with respect to $\xmin$, $\xmax$, to obtain the saddle point 
eigenvalue distribution $\rho(x)$, $y^I(x)$.

We shall be interested in evaluating the Wilson loop VEV (\ref{W}) in the fundamental 
representation, so that the Wilson loop is proportional to 
$\sum_{I=1}^\gauge\sum_{i=1}^N 
\ex^{\lambda_i^I}$. In the large $N$ limit, described by  the saddle point
density $\rho(x)$ and imaginary 
parts $y^I(x)$ of the eigenvalues, the VEV reduces simply to
\bea\label{Wmm}
\langle W\rangle_{\mathrm{QFT}} & =& N\sum_{I=1}^\gauge\int_{\xmin}^{\xmax} \ex^{x N^{{1}/{2}} + \ii y^I(x)}\rho(x)\dd x~.
\eea
Because of the form of $F[\rho(x),y^I(x)]$ for $\mathcal{N}=2$  Chern-Simons-matter theories, the saddle point eigenvalue density $\rho(x)$ is always a continuous, piecewise linear function on $(\xmin,\xmax)$. 
A simple computation then shows that, to leading order in the large $N$ limit, the  matrix model VEV (\ref{Wmm}) reduces to
\bea\label{WQFT}
\log{\langle W\rangle}_{\mathrm{QFT}} & = & \xmax\, N^{1/2}~.
\eea
This is our final formula for the large $N$ limit of the Wilson loop VEV. 
We see that it computes the 
\emph{maximum} value of the (real part of the) saddle point eigenvalues.

In our summary above we have suppressed the dependence on the R-charges
$\Delta_\alpha$ of the matter multiplets, labelled by $\alpha$, appearing in (\ref{oneloop}). 
If these are left arbitrary, one obtains a free energy $F$ that 
is a function of $\Delta_\alpha$, and according to \cite{Jafferis:2010un} the superconformal 
R-symmetry of an $\mathcal{N}=2$ superconformal field theory further extremizes 
$F$ as a function of $\Delta_\alpha$ (in fact maximizing $F$ \cite{Closset:2012vg}). For theories with M-theory duals of the form 
AdS$_4\times Y_7$ one finds the expected supergravity result
\bea\label{free}
F &=& \sqrt{\frac{2\pi^6}{27\, \Vol_\eta(Y_7)}}N^{3/2}~,
\eea
but as a \emph{function} of R-charges $\Delta_\alpha$ \cite{Martelli:2011qj}, where on the right hand 
side it is in general the \emph{contact volume} (\ref{contactvol}) of $Y_7$ that appears, 
as a \emph{function} of the Reeb vector field $\xi$. This has 
by now been demonstrated in many classes of examples in the literature 
 \cite{Gabella:2012rc}, \cite{Herzog:2010hf}--\cite{Amariti:2012tj}.

\section{BPS M2-branes}\label{sec:M2}

In this section we analyse the supersymmetric probe M2-branes that are relevant for computing 
the holographic dual of the Wilson loop VEV (\ref{WQFT}). We first recast the condition of supersymmetry into a 
geometric condition,  then derive the formula (\ref{M2action}) for the action of the M2-brane, and 
finally describe how this may be computed in practice using different geometric methods.

\subsection{Supergravity backgrounds}\label{sec:gravity}

We will study the general class of $\mathcal{N}=2$ supersymmetric AdS$_4\times Y_7$ backgrounds of M-theory
described in \cite{Gabella:2011sg, Gabella:2012rc}. We begin by recalling some relevant results and formulae. 
The eleven-dimensional metric and four-form $G_4$ take the form
\bea\label{11d}
g_{11} &=& \ex^{2\Delta}\left(\frac{1}{4}g_{\mathrm{AdS}_4} + g_{Y_7}\right)~,\nn\\
G_4 &=& \frac{m}{16}\mathrm{vol}_4 + F_4~,
\eea
where the metric on AdS$_4$ here has unit AdS radius, with volume form $\mathrm{vol}_4$. The warp factor $\Delta$ is taken to be a 
function on $Y_7$, $m$ is a constant, and $F_4$ is a four-form on $Y_7$. This is the most general ansatz 
compatible with the symmetries of AdS$_4$.
The eleven-dimensional Majorana spinor
takes the form
\bea\label{11dspinor}
\epsilon &=& \ex^{\Delta/2}\psi_+\otimes \chi_+ + \ex^{\Delta/2}\psi_-\otimes \chi_- + \mbox{charge conjugate}~,
\eea
where $\chi_\pm$ are complex spinors on $Y_7$, $\psi_\pm$ 
are the usual Killing spinors on AdS$_4$ (the $\pm$ signs 
are related to the charge under the R-symmetry, discussed below), 
and the factors of $\ex^{\Delta/2}$ have been introduced for convenience.

In general the spinors $\chi_\pm$ solve quite a complicated system 
of coupled first order equations on $Y_7$, that may be found in  \cite{Gabella:2011sg, Gabella:2012rc}. 
These equations are then necessary and sufficient for supersymmetry of the 
AdS$_4\times Y_7$ background. For our purposes we need  note only a few key formulae. 
We first define the real one-forms 
\bea\label{contact}
\xi &\equiv & \ii \bar\chi_+^c\gamma_{(1)}\chi_-~, \qquad \ssigma \ \equiv \ -\frac{6}{m}\ex^{3\Delta}\bar\chi_+\gamma_{(1)}\chi_+~,
\eea
where in general we denote $\gamma_{(n)}\equiv \frac{1}{n!}\gamma_{m_1\cdots m_n}\diff y^{m_1}\wedge\cdots \wedge \diff y^{m_n}$, 
with $y^1,\ldots, y^7$ local coordinates on $Y_7$,
and the superscript $c$ on the spinors denotes charge conjugation. 
By an abuse of notation, we'll more generally regard $\xi$ as the dual vector field defined by the metric $g_{Y_7}$. 
We then note that the differential equations for $\chi_\pm$ imply the equations
\bea\label{claude}
\bar\chi_+\chi_+ & = &  \bar\chi_-\chi_- \ = \ 1~,  \quad \frac{m}{6}\ex^{-3\Delta} \ = \ -\Imag \left[\bar\chi_+^c\chi_-\right]~, \quad \Real\left[\bar\chi_+^c\chi_-\right] \ = \ 0~,\nonumber\\
\Real\left[\bar\chi_+^c\gamma_{(1)}\chi_-\right] &=& 0~, \quad \bar{\chi}_+\gamma_{(1)}\chi_+ \ = \ - \bar{\chi}_-\gamma_{(1)}\chi_-~,\nonumber\\
\diff \ssigma & = & -\frac{12}{m}\ex^{3\Delta}\Real\left[\bar\chi_+^c\gamma_{(2)}\chi_-\right]~.
\eea
These equations may all be found in reference \cite{Gabella:2012rc}.

The one-form $\ssigma$ is a \emph{contact form} on $Y_7$, meaning that 
the top form $\ssigma\wedge (\diff\ssigma)^3$ is nowhere zero. Indeed, 
one finds \cite{Gabella:2012rc} that 
\bea\label{contactvolume}
\ssigma\wedge (\diff\ssigma)^3 &=& \frac{2^73^4}{m^3}\ex^{9\Delta}\vol_{7}~,
\eea
where $\vol_7$ is the Riemannian volume form defined by $g_{Y_7}$. It is a general 
fact that a contact form $\ssigma$ has associated to it a unique 
\emph{Reeb vector field}, defined by the relations
\bea\label{reebidentity}
\xi\lrcorner \ssigma &=& 1~, \qquad \xi\lrcorner \diff\ssigma \ = \ 0~,
\eea
and remarkably one finds that $\xi$ and $\ssigma$ defined by (\ref{contact}) indeed satisfy these equations. 
Moreover, $\xi$ is a Killing vector field under which $\chi_\pm$ carry charges $\pm 2$, and as 
such is the expected R-symmetry vector field.

Dirac quantization in this background implies that
\bea\label{dirac}
N &=& -\frac{1}{(2\pi \ell_p)^6}\int_{Y_7} *_{11} G_4 + \frac{1}{2}C_3\wedge G_4
\eea
should be an integer, where $\ell_p$ denotes the eleven-dimensional Planck length and $G_4=\diff C_3$. This may be identified with the 
M2-brane charge of the background, and 
a computation  \cite{Gabella:2011sg, Gabella:2012rc} gives
\bea\label{Nintegral}
N &=& \frac{1}{(2\pi\ell_p)^6}\frac{m^2}{2^53^2}\int_{Y_7}\ssigma\wedge (\diff \ssigma)^3~,
\eea
relating the quantized M2-brane charge to the contact volume (\ref{contactvol}) of $Y_7$ and  $m$. 
Since this is proportional to $m^2$, in fact the contact form in (\ref{contact}) may be defined 
only when this charge is non-zero, so that $m\neq 0$. We assume this henceforth.

The above supergravity solution of M-theory is valid only in the large $N$ limit, even for 
solutions with non-trivial warp factor $\Delta$ and internal four-form flux $F_4$. 
To see this \cite{Gabella:2012rc}, note that the scaling symmetry of eleven-dimensional supergravity in which the metric $g_{11}$ and 
four-form $G_4$ have weights two and three, respectively, leads to a symmetry in which 
one shifts $\Delta\rightarrow \Delta+\kappa$ and simultaneously scales $m\rightarrow \ex^{3\kappa}m$, 
$F_4\rightarrow \ex^{3\kappa}F$, where $\kappa$ is any real constant. 
We may then take the metric $g_{Y_7}$ on 
$Y_7$ to be of order $\mathcal{O}(1)$ in $N$, and conclude from the quantization condition 
(\ref{Nintegral}), which has weight 6 on the right hand side, and the expression for $m\ex^{-3\Delta}$ in (\ref{claude}) that $\ex^\Delta=\mathcal{O}(N^{1/6})$. 
It follows that the AdS$_4$ radius, while dependent on $Y_7$ in general, is $R_{\mathrm{AdS}_4}=\ex^\Delta=\mathcal{O}(N^{1/6})$, 
and that the supergravity approximation we have been using is valid only in the $N\rightarrow\infty$ limit.

\subsection{Choice of M-theory circle}\label{sec:Mcircle}

In addition to this background we must also pick a choice of M-theory circle. Geometrically, this 
means we also choose a $U(1)=U(1)_M$ action on $Y_7$. 
At first sight it might seem to be contradictory that the supergravity 
computation we describe
then manifestly depends on a choice of M-theory circle, while the dual superconformal field theory 
apparently does not. However, recall that the UV description of the gauge theory, whose Lagrangian we used 
to compute the localized path integral and Wilson loop in section \ref{sec:localize}, \emph{does} in fact require a choice 
of M-theory circle $U(1)_M$. We may have two or more such theories, arising from different choices of $U(1)_M$ and flowing to the same superconformal fixed point; but  it does \emph{not} follow that the Wilson loop operators in these theories map to each other. One thus expects the 
Wilson loop VEV to depend on a choice of M-theory circle, in general.

In terms of the supergravity solution described in the previous section, a choice of $U(1)_M$ 
implies the choice of a (non-$U(1)_R$) Killing vector field $\zeta_M$ on $(Y_7,g_{Y_7})$, whose 
flow generates the M-theory circle action. In particular $\zeta_M$ should preserve the Killing spinors 
$\chi_\pm$ on $Y_7$, and hence also the contact one-form 
$\ssigma$. The type IIA spacetime is then a warped product AdS$_4\times M_6$, where 
$M_6\equiv Y_7/U(1)_M$ is the quotient space.

Of course globally we must be careful when writing $M_6=Y_7/U(1)_M$.
Although in principle one might choose any $U(1)_M$ action on $Y_7$, in practice the gauge theories 
we study arise from ``nice'' actions of $U(1)_M$. In particular, if the action is \emph{free} then $M_6$ inherits 
the structure of a smooth manifold from $Y_7$, the simplest example being that of the ABJM theory 
with $M_6=\mathbb{CP}^3=S^7/U(1)_{\mathrm{Hopf}}$. If one embeds $S^7\subset \C^4$ 
as a unit sphere in the obvious way, then recall that $U(1)_{\mathrm{Hopf}}$ may be taken to have weights 
$(1,1,-1,-1)$ on the four complex coordinates $(z_1,z_2,z_3,z_4)$ on $\C^4$. In this 
case the dual field theory is the $\mathcal{N}=6$ ABJM theory, which in $\mathcal{N}=2$ language is a $U(N)\times U(N)$ 
Chern-Simons gauge theory with two chiral matter fields $A_1$, $A_2$ in the bifundamental $(\mathbf{N},\overline{\mathbf{N}})$ 
representation of this gauge group, two chiral matter fields $B_1$, $B_2$ in the conjugate $(\overline{\mathbf{N}}, \mathbf{N})$ 
representation, and a quartic superpotential. 

 Another important 
case is when $U(1)_M$ acts on $Y_7$ with a codimension four fixed point set $\FF\subset Y_7$, and is free 
on the complement of this fixed point set. In this case the action on the normal space $\R^4$ to a fixed point
is via $(w_1,w_2)\mapsto (\ex^{\ii\varphi}w_1,\ex^{\ii\varphi}w_2)$, where locally 
$\zeta_M=\partial_\varphi$ and $(w_1,w_2)$ are complex coordinates on $\R^4=\C\oplus \C$. 
In this case the quotient normal space is $\R^3=\R^4/U(1)_M$, with the fixed point set $\FF$ at the origin becoming a D6-brane 
locus in the type IIA spacetime. With this understanding, the IIA spacetime is again a smooth AdS$_4\times M_6$, but 
with an AdS$_4$-filling D6-brane wrapping $\FF$, now thought of as a submanifold $\FF\subset M_6$.
 Again, the simplest example is a quotient of $S^7$, but 
now where $U(1)_M$ has weights $(1,-1,0,0)$ on the coordinates $(z_1,z_2,z_3,z_4)$ on $\C^4\supset S^7$. 
This fixes a copy of $\FF=S^3\subset S^7$ at $z_1=z_2=0$, which then becomes a D6-brane locus 
in the type IIA spacetime $M_6=S^7/U(1)_M=S^6$. The dual field theory is then the low-energy gauge theory on $N$ D2-branes in flat spacetime, which in $\mathcal{N}=2$ language is 
a $U(N)$ gauge theory with adjoint fields $X_1, X_2, X_3$ and cubic superpotential ($\mathcal{N}=8$ super-Yang-Mills), but with 
additional fundamental fields arising from the low-energy string modes stretching between the
D2-branes and D6-brane. This gives rise to additional fields $q$, $\tilde{q}$ in the fundamental and anti-fundamental 
of $U(N)$ respectively, and a corresponding additional superpotential term (see \cite{Benini:2009qs}). This is often called the mirror 
to the ABJM theory, and indeed \emph{both} theories have superconformal fixed points 
that are dual to AdS$_4\times S^7$. The gauge theories are of course quite different, 
one being a $U(N)\times U(N)$ gauge theory, the other being a $U(N)$ gauge theory.\footnote{It happens that the Wilson loops turn out to be the same in these theories (essentially due to the 
high degree of symmetry), but the spectrum of BPS M2-branes/fundamental strings is certainly different. 
See section \ref{sec:S7}}

In the above cases the type IIA description is under control and typically 
well-understood, allowing one to determine an appropriate  UV gauge theory. 
We shall see more complicated examples in section \ref{sec:examples}.

\subsection{BPS M2-brane probes}

The supersymmetric M2-brane which is conjectured to be holographically dual to the Wilson loop on $S^3$ must necessarily have as boundary a Hopf circle in $S^3$. A convenient explicit form for the Euclidean  AdS$_4$ metric can be taken to be
\bea
g_{\mathrm{AdS}_4} &  = & \frac{\dd q^2}{1+q^2}+q^2 \dd\Omega_3~,
\eea
with $\dd\Omega_3$ the round metric on the unit sphere $S^3$, and $q\in[0,\infty)$ a radial coordinate. The M2-branes of interest then wrap $\Sigma_2\times S^1_M$, where the surface $\Sigma_2\subset \mathrm{AdS_4}$ has boundary $\partial \Sigma_2=S^1_{\mathrm{Hopf}}\subset S^3$, and $ S^1_M\subset Y_7$ is the M-theory circle. The submanifold $\Sigma_2$ is then parametrized by the radial direction $q$ in $\mathrm{AdS_4}$, and a geodesic Hopf circle $S^1_{\mathrm{Hopf}}$ in $S^3$, whilst $ S^1_M\subset Y_7$ is {\it a priori} arbitrary (imposing supersymmetry will later give restrictions on $ S^1_M$). The area of the surface $\Sigma_2$ in $\mathrm{AdS_4}$ is divergent, but can be regularized by subtracting the length of its boundary, {\it i.e.} the length of the $S^1_{\mathrm{Hopf}}$ geodesic in $S^3$ at $q\to\infty$. Notice this 
is then a local boundary counterterm. Including also the warp factor one finds the regularized area to be
\bea
\Vol(\Sigma_2) & =& -\frac{\pi}{2}\ex^{2\Delta}~.
\eea
The action of the M2-brane then reads
\bea\label{actionM2prov}
S_{\mathrm{M2}}& =& \frac{\Vol(\Sigma_2\times S^1_M)}{(2\pi)^2\ell_p^3}\ = \ -\frac{1}{(2\pi)^2\ell_p^3}\frac{\pi}{2}\int_{ S^1_M}\ex^{3\Delta}\vol_{ S^1_M}~,
\eea
where $\vol_{ S^1_M}$ is the volume form on $ S^1_M$ induced from the metric $g_{Y_7}$.

As mentioned above, imposing that the M2-brane $\Sigma_2\times S^1_M$ is supersymmetric gives restrictions on the possible circles $ S^1_M$. To see this, we need to split the Clifford algebra $\mathrm{Cliff}(11,0)$ generated by gamma matrices $\Gamma_A$ satisfying $\{\Gamma_A,\Gamma_B\}=2\delta_{AB}$ into $\mathrm{Cliff}(4,0)\otimes\mathrm{Cliff}(7,0)$ via
\be
\Gamma_{\alpha} \ = \ \rho_\alpha\otimes 1~,\qquad \Gamma_{a+3}\ =\ \rho_5\otimes\gamma_a~,
\ee
where $\alpha,\beta=0,1,2,3$ and $a,b=1,\ldots,7$ are orthonormal frame indices for Euclidean $\mathrm{AdS_{4}}$ and $Y_7$ respectively, $\{\rho_\alpha,\rho_\beta\}=2\delta_{\alpha\beta}$, $\{\gamma_a,\gamma_b\}=2\delta_{ab}$ and we have defined $\rho_5\equiv \rho_0\rho_1\rho_2\rho_3$. If we denote by $X^M$ the embedding coordinates of the worldvolume of the M2-brane into the target geometry, the amount of preserved supersymmetry is equal to the number of spinors $\e$, as in \eqref{11dspinor}, satisfying the projection condition \cite{Becker21995}
\be
\mathbb{P}\e\ =\ 0~,\quad \mbox{where}\quad \mathbb{P} \ \equiv \ \frac{1}{2}\left(1-\frac{\ii}{3!}\eps^{ijk}\de_iX^M\de_jX^N\de_kX^P\Gamma_{MNP}\right)~,
\ee
with $i,j,k$ indices on the worldvolume. We now choose an orthonormal frame in eleven-dimensions 
as ({\it c.f.} (\ref{11d}))
\bea
E^0 &=& \frac{1}{2}\ex^{\Delta}\frac{\diff q}{\sqrt{1+q^2}}~, \quad E^m \ = \  \frac{1}{2}\ex^{\Delta} q e^m~, \quad E^{3+a}  \ = \  
\ex^{\Delta}e^a_{Y_7}~,
\eea
where $\{e^m\}_{m=1,2,3}$ is an orthonormal frame on $S^3$ and $\{e^a_{Y_7}\}_{a=1,\ldots 7}$ is 
an orthonormal frame on $(Y_7,g_7)$, with $e^1_{Y_7}$ (or rather its dual vector field) aligned along the M-theory circle vector field $\zeta_M$. 
Taking $e^3$ to be aligned along the Hopf circle, as in section \ref{sec:WL}, 
the projector $\mathbb{P}$ then takes the simple form
\bea
\mathbb{P} &= &\frac{1}{2}\left(1-\ii\rho_5\rho_{ 0 3}\otimes\gamma_{ 1}\right)~,
\eea
and the constraints that follow on the spinors $\psi_\pm$, $\chi_\pm$ on Euclidean AdS$_4$ and $Y_7$, respectively, are
\be\label{projfact}
(1-\ii\rho_5\rho_{0 3})\psi_{\pm} \ = \ 0~, \quad\mathrm{and}\quad (1-\gamma_{ 1})\chi_\pm \ = \ 0~.
\ee

In order to determine how much supersymmetry is preserved by the brane in $\mathrm{AdS_4}$, we must count the number of Killing spinors $\psi_{\pm}$ that satisfy the last projection equation. We may decompose the four-dimensional gamma matrices into $\rho_0=1\otimes\tau_3$ and $\rho_\mu=\tau_\mu\otimes \tau_1$, with the Pauli matrices $\tau_\mu$, $\mu=1,2,3$. These matrices act on spinors of the form $\psi=(\psi_1, \psi_2)^T$, with $\psi_{1,2}$ 2-component spinors. The Killing spinors on AdS$_4$ may then be constructed from 
Killing spinors on the $S^3$ at fixed radial coordinate $q$. Explicitly, if $\et$ solves the Killing spinor equation
\bea\label{Kill}
\nabla_\mu\et  &= & \frac{\ii}{2}\tau_\mu\et~,
\eea
on $S^3$, then 
\be
\psi \ = \ \begin{pmatrix}(q+\sqrt{1+q^2})^{1/2}\et\\ (q+\sqrt{1+q^2})^{-1/2}\et\end{pmatrix}~,
\ee
is a Killing spinor on Euclidean AdS$_4$. Equation (\ref{Kill}) has two solutions, one being chiral and one anti-chiral, {\it i.e.} $\tau_3\et=\pm\et$. 
One then easily shows that the first projection equation in \eqref{projfact} is satisfied if we restrict to chiral $\et$ in the last solution, which singles out one of these two spinors on AdS$_4$.\footnote{The other two Killing spinors on AdS$_4$ are constructed from spinors on $S^3$ satisfying $\nabla_\mu\et  = -\frac{\ii}{2}\tau_\mu\et$. 
We set the corresponding spinors to zero in section \ref{sec:Wilson}, as they are not used in the supersymmetric localization. 
Again, one chirality is  broken by the M2-brane.} Hence the M2-brane preserves half of the supersymmetry in $\mathrm{AdS_4}$. Note that the same positive chirality condition also appeared 
in the supersymmetry condition derived in the field theory context, {\it c.f.} \eqref{Wsusycondition}.\footnote{Notice 
that the Wilson loop circle $\gamma\subset S^3$ is calibrated by $e^3$, one of the left-invariant one-forms under $SU(2)_{\mathrm{left}}$, which is 
a contact form on $S^3$.}

The second projection equation in \eqref{projfact}  tells us which circles $ S^1_M$ give rise to supersymmetry-preserving M2-branes. Following a standard argument one notices that
\be
\bar\chi_+\left(\frac{1-\gamma_{ 1}}2{}\right)\chi_+ \ = \ \bar\chi_+\left(\frac{1-\gamma_{1}}2{}\right)^\dag\left(\frac{1-\gamma_{ 1}}2{}\right)\chi_+ \ = \ \left|\left(\frac{1-\gamma_{1}}2{}\right)\chi_+\right|^2 \ \geq \ 0~,
\ee
using $\gamma_{1}=\gamma_{ 1}^\dag$ and $\gamma_{ 1}^2=1$. This immediately gives $\vol_{ S^1_M}\geq\bar\chi_+\gamma_{(1)}\chi_+$ 
(with a pull-back understood), with equality if and only if some supersymmetry is preserved by $ S^1_M$. The action \eqref{actionM2prov} for a supersymmetric brane is then
\be
S_{\mathrm{M2}} \ = \ \frac{\Vol(\Sigma_2\times S^1_M)}{(2\pi)^2\ell_p^3} \ =  \ -\frac{1}{(2\pi)^2\ell_p^3}\frac{\pi}{2}\int_{S^1_M}\ex^{3\Delta}\bar\chi_+\gamma_{(1)}\chi_+~.
\ee
With the help of equations \eqref{contact} and \eqref{Nintegral} the action of a supersymmetric M2-brane can be rewritten in terms of the contact form $\ssigma$ as (taking a convention in which $m<0$)
\bea\label{M2actionagain}
S_{\mathrm{M2}} & =& -\frac{(2\pi)^2\int_{S^1_M}\ssigma}{\sqrt{2\int_{Y_7}\ssigma\wedge (\diff\ssigma)^3}}N^{1/2}~.
\eea

\subsection{M-theory Hamiltonian function}

In this subsection we further elucidate the geometry associated to these supersymmetric M2-branes. 
This geometric structure will both be of practical use, when we come to compute 
the M2-brane actions (\ref{M2actionagain}) in examples, and also, as we will see, is realized rather directly in the large $N$ dual matrix model.

We begin by introducing the \emph{M-theory Hamiltonian function}
\bea
h_M &\equiv & \ssigma(\zeta_M) \ = \ \zeta_M\lrcorner \ssigma~,
\eea
where $\zeta_M$ generates the M-theory circle action.
This is a real function on $Y_7$, and since $\zeta_M$ is assumed to preserve the Killing spinors and metric on $Y_7$, 
it follows that $\zeta_M$ preserves $h_M$ and commutes with the Reeb vector field $\xi$. 
It follows that the contact length of an M-theory circle $S^1_M$ over a point $p\in M_6=Y_7/U(1)_M$ is
given by $\int_{S^1_M}\ssigma = 2\pi h_M(\hat{p})$, where $\hat{p}\in Y_7$ is any lift of the point $p$.
This directly leads to the form of the M2-brane action (\ref{M2action}).

One way to characterize the \emph{supersymmetric} M-theory circles $S^1_M$ is to note that 
on  $TY_7\mid_{S^1_M}$ the vector $\zeta_M$ is necessarily 
 proportional to the Reeb vector. Indeed, using \eqref{claude} one can show that at these supersymmetric points
\bea\label{zetaReeb}
\zeta_M\lrcorner\dd\ssigma & =& 0~.
\eea
To see this one takes the projection condition (\ref{projfact}) with $\chi_-$,  applies $\bar{\chi}_+^c\gamma_a$ on the left, and 
then takes the real part of the resulting equation. Using $\Real [\bar{\chi}_+^c\chi_-] = \Real [\bar{\chi}_+^c\gamma_a\chi_-]=0$ 
and the relation between $\diff\ssigma$ and $\Real [\bar{\chi}_+^c\gamma_{(2)}\chi_-]$ in (\ref{claude}) then leads to (\ref{zetaReeb}). 
That this then implies $\zeta_M\propto \xi$ follows from the fact that $\ssigma$ is a contact form: $\diff\ssigma$ is 
a symplectic form on $\ker \ssigma$, the rank 6 subbundle of the tangent bundle $TY_7$ of $Y_7$ 
defined as vectors having zero contraction with $\ssigma$.
Since this means that $\diff\ssigma$ is non-degenerate on this rank 6 bundle, 
and since also $TY_7=\ker\ssigma\oplus \langle\xi\rangle$, where $ \langle\xi\rangle$ is the real line bundle 
 spanned by vectors proportional to $\xi$,
 (\ref{zetaReeb}) implies that the projection of $\zeta_M$ onto  $\ker \ssigma$ is zero, {\it i.e.} that
 $\zeta_M\propto\xi$.

The condition (\ref{zetaReeb}) is then also the condition that we are at a \emph{critical point} of the Hamiltonian $h_M$. 
To see this, we rewrite $\mathcal{L}_{\zeta_M}\ssigma=0$ using the Cartan formula, so that (\ref{zetaReeb}) is equivalent to
\bea
\diff(\zeta_M\lrcorner\ssigma) &=& 0 \qquad \Leftrightarrow \qquad \diff h_M \ = \ 0~.
\eea
Thus the supersymmetric M2-branes lie precisely on the critical set $\{\diff h_M=0\}$, and their 
action (\ref{M2action}) is determined by $h_M$ evaluated at the critical point! It is a general 
fact that any component of the moment map for a compact group action on a symplectic manifold is a 
\emph{Morse-Bott function}. Here more precisely recall that the cone 
$C(Y)=\R_{\geq 0}\times Y_7$ is \emph{symplectic}, with symplectic form
\bea
\omega &=& \frac{1}{2}\diff\left( r^2\ssigma\right)~,
\eea
where $r\geq 0$ is a radial coordinate.
In fact the cone being symplectic is equivalent to $(Y_7,\ssigma)$ being contact. The M-theory circle action
then gives a $U(1)_M$ action on this cone, with moment map
\bea
\mu &=& \frac{1}{2}r^2\zeta_M\lrcorner \ssigma~.
\eea
Thus $\mu$ is Morse-Bott, and the restriction of $\mu$ to $Y_7$ at $r=1$ is 
our Hamiltonian function $h_M/2$. We thus know that 
the image $h_M(Y_7)=[\cmin,\cmax]$ is a closed interval, 
and this is further subdivided into $\n$ intervals via 
$\cmin=c_1<c_2<\cdots<c_{\n+1}=\cmax$, where the $c_i$ are images 
under $h_M$ of the critical set $\{\diff h_M=0\}$.
On each open interval $c\in(c_i,c_{i+1})$ the level surfaces 
$h_M^{-1}(c)$ are all diffeomorphic to the same fixed six-manifold, 
with the topology changing as one crosses a critical point $c_i$. 

Finally, since at a supersymmetric $S^1_M$ we have $\zeta_M\propto \xi$, it follows that 
the corresponding point $p\in M_6=Y_7/U(1)_M$ is a \emph{fixed point} 
under the induced Reeb vector action on $M_6=Y_7/U(1)_M$. 
That is, over every fixed point $p\in M_6$ of $\xi$,  there exists a calibrated and supersymmetric M-theory circle $S^1_{M,p}$ whose corresponding supersymmetric M2-brane action is given by (\ref{M2action}).

In the holographic computation of the Wilson loop VEV via the M2-brane action, one should \emph{sum} $\ex^{-S_{\mathrm{M2},p}}$  over all contributions. 
In some cases 
(typically with more symmetry) we shall find that the supersymmetric points $p\in M_6$ form \emph{submanifolds} which are fixed by $\xi$, 
and this sum in fact becomes an integral over the different connected submanifolds. Notice that 
$h_M$ is constant on each connected component of the fixed point set.
In any case,
in the large $N$ limit only the \emph{longest} circle $S^1_M$ survives, the others being exponentially suppressed relative to it in the 
sum/integral, hence proving formula \eqref{Wgravity}.

The calculation of the action of a supersymmetric M2-brane can be completely carried out once the Reeb vector field  $\xi$ and the M-theory circle generator $\zeta_M$ are known. Indeed, the contact volume $\Vol_\ssigma(Y_7)$ is a function only of the  Reeb vector \cite{Gabella:2010cy}, and the length of a calibrated circle $\int_{S^1_{M,p}}\ssigma=2\pi h_M(\hat{p})$ depends only on $\xi$, $\zeta_M$ and the point $p$. Even though this could appear to be involved, the computation of these two quantities is relatively straightforward for appropriate classes of $Y_7$. In particular, if we focus on toric  manifolds, standard geometrical techniques can be exploited to straightforwardly find \emph{all} calibrated circles, {\it i.e.} the connected 
components of the critical set 
$\{\diff h_M=0\}\subset Y_7$, as well as the contact volume \cite{Martelli:2006yb}. This is the subject of the next subsection. 

\subsection{Geometric methods of computation}\label{sec:geometry}

In this section we explain how to compute the various quantities we have been discussing in 
appropriate classes of examples. We focus our discussion on \emph{toric} geometries, which means 
that $U(1)^4$ acts on $Y_7$, preserving the contact form $\ssigma$. In this case there are 
 some pretty geometric methods, first developed in \cite{Martelli:2005tp, Martelli:2006yb}, that may be utilized to calculate the length of the calibrated M-theory circles, as well as the volumes of the internal spaces.  We will thus focus on this class of solutions, although we note 
 that the more general methods described in \cite{Martelli:2006yb} may be used to attack non-toric cases.

Let us begin with the \emph{symplectic cone} $(C(Y)=\R_{\geq 0}\times Y, \omega=\frac{1}{2}\diff (r^2\ssigma))$, 
but in general dimension $2n$. Equivalently, $(Y,\ssigma)$ is contact with $\dim Y=2n-1$. 
The toric condition means that 
$U(1)^n$ acts on the symplectic cone $C(Y)$ preserving the symplectic form $\omega$, and we may parametrize the generating vector fields as
$\partial_{\phi_i}$, with $\phi_i\in[0,2\pi)$ and $i=1,\ldots,n$. 
 This allows one to introduce symplectic coordinates $(y_i,\phi_i)$ in which the symplectic form on $C(Y)$ has the simple expression 
\bea
\omega \ = \ \sum_{i=1}^n\dd y_i\wedge\dd\phi_i~.
\eea
Moreover, when the toric cone is of \emph{Reeb type}, meaning that $\xi$ is in the Lie algebra 
 of $U(1)^n$, the coordinates $y_i$ take values in a convex polyhedral cone $\mathcal{C}^*\subset \mathbb{R}^n$ \cite{lerman}. If this cone has $d$ facets, we have corresponding
outward primitive normal vectors to these facets, $v_a\in \Z^n$, $a=1,\ldots,d$, with the facets corresponding (under the moment map) to the fixed 
point sets  of $U(1)\subset U(1)^n$ with weights $v_a$. In particular this set-up applies to toric
Sasakian $Y$ \cite{Martelli:2005tp}, in which the symplectic cone $C(Y)$ is also \emph{K\"ahler}. 
In this case, the topological condition that $C(Y)$ is Calabi-Yau (more precisely, 
that the apex $\{r=0\}$ is a Gorenstein singularity) 
is equivalent 
to the existence of an $SL(n,\mathbb{Z})$ transformation such that the normal vectors take the form $v_a=(1,w_a)$, for all $a$, with $w_a\in\mathbb{Z}^{n-1}$. In this basis, the first component of the Reeb vector is necessarily $\xi_1=n$ \cite{Martelli:2005tp}.

In general the components of $\xi=\sum_{i=1}^n\xi_i\partial_{\phi_i}$ form  a vector $\vec\xi=(\xi_1,\ldots,\xi_n)$ that defines the \emph{characteristic hyperplane} in $\mathbb{R}^n$: $\{\vec y\in\mathbb{R}^n\mid \vec \xi\cdot\vec y=\frac{1}{2}\}$. This hyperplane interesects $\mathcal{C}^*$ to form a finite polytope $\Delta_\xi$, and the contact volume of the base $Y$ is related to the volume of this polytope by\footnote{
In the Sasakian case the Riemannian volume and contact volumes coincide.}
\be
\Vol_\ssigma(Y) \ = \ 2n(2\pi)^n\Vol(\Delta_\xi)~.
\ee
Moreover, each of the $d$ facets $\mathcal{F}_a$, intersected with the characteristic hyperplane,  are images under the moment map of  $(2n-3)$-dimensional subspaces $\Sigma_a$ of $Y$. The volumes of these submanifolds may be calculated once the volumes of the facets are known, for
\be
\Vol_\ssigma(\Sigma_a) \ = \ (2n-2)(2\pi)^{n-1}\frac{1}{|v_a|}\Vol(\mathcal{F}_a)~.
\ee
In addition, the volume of the base manifold $Y$ is simply given by 
\be\label{volY}
\Vol_\ssigma(Y) \ = \ \frac{(2\pi)^n}{\xi_1}\sum_{a=1}^d\frac{1}{|v_a|}\Vol(\mathcal{F}_a)~.
\ee

 In \cite{Martelli:2006yb} the idea is to study the space of K\"ahler cone metrics on $C(Y)$, and thus Sasakian structures on $Y$. One then 
considers the Einstein-Hilbert action (with a fixed positive cosmological constant) restricted to this space of Sasakian metrics on $Y$, 
so that a Sasaki-Einstein metric on $Y$ is a critical point. In fact the action is minimized and proportional to the volume of the base $\Vol(Y)$ when the metric on $Y$ is Sasaki-Einstein. In this case 
 there is \emph{unique} Reeb vector of the form $\vec\xi=(n,\xi_2,\ldots,\xi_n)$ such that the Einstein-Hilbert action, or equivalently $\Vol(Y)$, is minimized as a function of $\xi$. Thus, for any given toric diagram one calculates $\Vol(Y)$ with formula \eqref{volY} as a function of the Reeb vector, 
and determines $\vec\xi$ for the Sasaki-Einstein metric on $Y$ by minimizing this function.\footnote{That 
the Sasaki-Einstein metric indeed always exists was proven in \cite{Futaki:2006cc}.} Presumably these ideas 
extend to more general warped geometries, with non-zero internal flux $F_4\neq 0$ in (\ref{11d}), 
following a similar construction in type IIB AdS$_5$ solutions \cite{Gabella:2010cy}.

In this paper we need only apply this method for $n=4$. A way to compute $\Vol(\mathcal{F}_a)$ as a function of the Reeb vector for $n=4$ has been described in \cite{Franco:2008um}. If the facet $\mathcal{F}_a$ is a tetrahedron, its vertex is at the origin in $\mathcal{C}^*$ and its base is a triangle lying in the characteristic hyperplane. This is generated by three edges 
passing from the characteristic hyperplane to the origin, and bounded by four hyperplanes creating the polyhedron. In addition to $v_a$, three other facets are then involved in the construction of the tetrahedron, and we denote their normal vectors as $v_{a,1}, v_{a,2}, v_{a,3}$. The volume of the tetrahedron 
may be expressed as 
\be
\frac{1}{|v_a|}\Vol(\mathcal{F}_a)\ = \ \frac{1}{48}\frac{(v_a,v_{a,1},v_{a,2},v_{a,3})^2}{|(\xi,v_{a},v_{a,1},v_{a,2})(\xi,v_{a},v_{a,1},v_{a,3})(\xi,v_{a},v_{a,2},v_{a,3})|}~,
\ee
with $(\cdot,\cdot,\cdot,\cdot)$ the determinant of a $4\times 4$ matrix. If the facet $\mathcal{F}_a$ is not a tetrahedron, {\it i.e.} there are more than 3 edges that meet at a vertex in the toric diagram ({\it c.f.} below), the volume can be computed with the same formula by breaking up the facet into tetrahedrons. 

The \emph{toric diagram} for a toric Calabi-Yau cone is by definition the convex hull of the lattice vectors $w_a$ in $n-1=3$ dimensions. To each vertex in this diagram corresponds a facet $\mathcal{F}_a$. If the vertex is located at the intersection of three planes, or equivalently three edges of the toric diagram meet at the vertex, then it corresponds to a tetrahedron. If instead four edges meet at the vertex, the facet is a pyramid that can be split into two tetrahedrons, and so on. A given facet $\mathcal{F}_a$ then corresponds to a vector $v_a=(1,w_a)$, with $w_a$ a vertex in the toric diagram; the other three vectors $v_{a,1}, v_{a,2}, v_{a,3}$ are the outward-pointing primitive vectors corresponding in the toric diagram to the three edges 
that meet at the vertex $v_a$. Let us also note that the base $Y_7$ of the cone is a \emph{smooth manifold} 
only if each face of the toric diagram is a triangle, and there are no lattice points internal to any edge or face. 
These conditions are equivalent \cite{Benishti:2009ky} to the cone being \emph{good}, in the sense of \cite{lerman}. 

It should now be clear that once a toric diagram is given for a toric Calabi-Yau cone $C(Y)$, one can calculate the volume of the base $\Vol_\ssigma(Y)$ as a function of the toric data and the Reeb vector that is parametrized by $\vec\xi=(4,\xi_2,\xi_3,\xi_4)$. After minimizing the volume with respect to $\xi_2,\xi_3,\xi_4$, one obtains the Reeb vector and $\Vol_\ssigma(Y)$ as a function of the toric data only. For more general 
warped solutions with flux the cone is not Ricci-flat K\"ahler, but in the examples we shall study later 
in section \ref{sec:examples} the Reeb vector $\vec\xi$ and toric contact structure are in fact known \cite{Gabella:2012rc}.

Next we turn to the M-theory Hamiltonian function $h_M$, and the computation of the calibrated circles in $Y_7$ and their lengths.
This involves, by definition, the 
choice of an M-theory circle $\zeta_M$, as described in section \ref{sec:Mcircle}. 
 As we proved in the last section, supersymmetric calibrated $S^1_M$ exist where $\zeta_M$ is parallel to $\xi$. 
This is equivalent to
\be
\zeta_M\ = \ \ssigma(\zeta_M)\xi \ = \ h_M\xi~,
\ee
as follows by taking the contraction of each side with $\ssigma$. 
We can conclude that if we know the proportionality constant between $\zeta_M$ and $\xi$,
the length of the corresponding calibrated M-theory circle, located over  a fixed point $p$ under $\xi$ in $M_6$, is then 
simply $2\pi h_M(\hat{p})$ with $\hat{p}\in Y_7$ any point projecting to $p$. In terms of the toric 
geometry above, notice that
\bea\label{Hamtoric}
h_M &=& 2\sum_{i=1}^n y_i\zeta_M^i~,
\eea
where $\zeta_M=\sum_{i=1}^n \zeta_M^i\partial_{\phi_i}$. This may be regarded as a function on the 
polytope $\Delta_\xi$, that is the image of $Y_7$ under the moment map.

The only remaining question is how to find where the two vectors $\zeta_M$, $\xi$ are proportional to each other, 
or equivalently what are the critical points of $h_M$, and also what is the value of $h_M$ at those points.
With the formalism at hand, this is straightforward to answer. Once a toric diagram and $\zeta_M$ are given, the Reeb vector (and the volume) can be found with the method described above. We may then find the solutions to the equation
\be\label{zetaxilin}
\zeta_M \ = \ \beta\xi+\sum_{a\in I}\alpha_a v_a~,
\ee
with $\beta,\alpha_a$ real numbers, and $I\subset \{1,\ldots,d\}$ a subset of three facets which \emph{intersect}. 
Geometrically, the intersection of three facets defines an edge of $\mathcal{C}^*$, which corresponds to a 
\emph{circle} $S^1\subset Y_7$. This circle is a fixed point set of $U(1)^3\subset U(1)^4$ defined by the three 
vectors $v_a$, $a\in I$, meaning that the generating $U(1)$ vector fields corresponding to $v_a$ 
are zero over this circle, and hence $\zeta_M$ is parallel to $\xi$. Thus this $S^1$ is precisely a
calibrated circle. The proportionality constant is then $h_M=\ssigma(\zeta_M)=\beta$, and  its length is $2\pi h_M$. 
Thus our problem boils down to linear algebra on the polyhedral cone. 

We make a few further geometrical observations. First, if (\ref{zetaxilin}) holds with $\beta= 0$ then 
$\zeta_M$ actually fixes the $S^1$, meaning that there must be D6-branes present. The M-theory circle then has zero length 
on such loci, formally leading to M2-branes with zero action; if $\zeta_M$ 
acts freely on $Y_7$ this cannot happen. 
Next we note that (\ref{zetaxilin})  cannot hold 
with $\alpha_a=0$ for \emph{all} $a\in I$, since then $\zeta_M$ would be \emph{everywhere} 
parallel to $\xi$, and this cannot happen since $\zeta_M$ is a non-R symmetry. However, 
it may happen that (\ref{zetaxilin}) holds with one or two (but not all three) of the coefficients $\alpha_a=0$. 
Geometrically, this means that in this case $\zeta_M$ is parallel to $\xi$ over the intersection of (respectively) 
the corresponding two or one facets with non-zero $\alpha_a$ coefficients, leading to 
three-dimensional or five-dimensional subspaces of $Y_7$ which are fibred by calibrated $S^1_M$ circles. 
These then descend to two-dimensional or four-dimensional fixed point sets of $\xi$ on $M_6=Y_7/U(1)_M$, 
respectively. We shall see examples of this in the next section. Finally, 
if the toric diagram contains faces which have more than three sides, then (\ref{zetaxilin}) 
may hold for $I$ being the corresponding set of 4 or more vectors $v_a$. 
In this case the manifold has a locus of (worse than orbifold) singularities 
along the corresponding $S^1$ in $Y_7$, and our theory above doesn't directly 
apply to these singular circles (their tangent spaces are not even a quotient of $\R^7$).

Even though the above theoretical background may appear cumbersome, it is effectively not difficult to find the volume of $Y_7$, its Reeb vector $\xi$ and all the calibrated circles and their lengths. Thanks to equation \eqref{M2action}, the action for each corresponding M2-brane follows straightforwardly, and can be compared to the data extracted from the matrix model of the dual field theory. We examine these computations, in a variety 
of examples, in the next section. 

\subsection{Hamiltonian function and density}
In \cite{Gulotta:2011si, Gulotta:2011aa} a relation was also found between $\rho(x)$, and other matrix model 
variables, and certain geometric invariants. In particular, equation (1.4a) of  \cite{Gulotta:2011aa} relates 
$\rho(x)$ to the derivative of a function that counts operators in the chiral ring of the gauge theory according 
to their R-charge and monopole charges. In the language of the current paper, the monopole charge is the charge under $U(1)_M$. With our notations and conventions, using \cite{Gulotta:2011aa} one can rewrite their conjecture for $\rho(x)$ in the following form:
\bea\label{rhohm}
\rho(x) & =& \left.\frac{4}{\pi^2}\frac{(2\pi)^3}{\sqrt{96\, \Vol_\ssigma(Y_7)}}\frac{\de_r\vol(P_{rc})}{|\xi\wedge\zeta_M|}\right|_{r=1}~,\nn\\   \mbox{where} \quad  P_{rc} & \equiv &  \left\{y\in\mathcal{C}^*\Big|\ \vec y\cdot\vec\xi\ =\ \frac{r}{2}\ ,\ \vec y\cdot\vec\zeta_M\ =\ \frac{c}{2}\right\}\ ,
\eea
where the variable $c$ is related to $x$ by \eqref{xc}. Using equation \eqref{Hamtoric}, we know that for the toric case $\vec y\cdot\vec\zeta_M=\frac{1}{2}h_M$. If we introduce 
\be
P_c\ \equiv \ \left\{y\in\mathcal{C}^*|\ h_M \ =\ c\right\}~,
\ee
we see that $P_{rc}$ is nothing but the intersection of $P_c$ with the characteristic hyperplane. But since the pre-image (under the moment map) of $P_c$ in $Y_7$ is the same as $h^{-1}_M(c)$, which changes topology every time we pass through a critical point of $h_M$, we know that the topology of the pre-image of $P_{rc}$ in $Y_7$ also changes every time a critical point is crossed. Thus we expect a change of behaviour of $\vol(P_{rc})$ and hence $\rho(x)$ at the critical points $x_i$ that are related to the $c_i$ by \eqref{xc}. In other words, the eigenvalue density has a different behaviour in each subset $(c_i,c_{i+1})$, as we will see in the examples in the next section, because there are supersymmetric M2 branes located at the $c_i$, which are critical points of a Hamiltonian function. That explains why the function $\rho(x)$ has a jump in its derivative precisely at the critical points.

\section{Examples}\label{sec:examples}

In this section we illustrate the duality between geometries and matrix models in a wide variety of examples. 
In particular we will compute the image of the M-theory Hamiltonian $h_M(Y_7)=[\cmin,\cmax]$, and 
show that it coincides with the support of the matrix model eigenvalues $[\xmin,\xmax]$ via (\ref{xc}). The critical points 
of $h_M$ will be shown to map to the points $x=x_i$ where $\rho'(x)$ has a jump discontinuity, with the 
matching of Wilson loops being a corollary of this result for $x=\xmax$. 

\subsection{Duals to the round $S^7$}\label{sec:S7}

We begin by studying two superconformal duals to AdS$_4\times S^7$, where $S^7$ is equipped with its standard round Einstein metric. 
These differ in the choice of M-theory circle $U(1)_M$ acting on $S^7$, as we discussed briefly in section \ref{sec:Mcircle}. In this case the geometry is particularly simple, 
allowing us to illustrate the geometric structures we have described very explicitly.

\subsubsection{ABJM theory}\label{sec:ABJM}

The ABJM theory \cite{ABJM} is an $\mathcal{N}=6$ superconformal $U(N)_k\times U(N)_{-k}$  
Chern-Simons-matter theory. In $\mathcal{N}=2$ language,
there are two chiral matter fields $A_1$, $A_2$ in the bifundamental $(\mathbf{N},\overline{\mathbf{N}})$ 
representation of this gauge group, two chiral matter fields $B_1$, $B_2$ in the conjugate $(\overline{\mathbf{N}}, \mathbf{N})$ 
representation, and a quartic superpotential. Here the subscript $k\in\Z$ in $U(N)_k$ denotes the Chern-Simons 
level for the particular copy of $U(N)$, as in (\ref{CS}). This theory is dual to AdS$_4\times S^7/\Z_k$ 
with $N$ units of flux (\ref{Nintegral}), where $\Z_k\subset U(1)_{\mathrm{Hopf}}=U(1)_M$. 

We may realize $S^7$ as the unit sphere $S^7\subset\R^8\cong \C^4$ and take $U(1)_{\mathrm{Hopf}}$  to have weights 
$(1,1,-1,-1)$ on the four complex coordinates $(z_1,z_2,z_3,z_4)$ on $\C^4$. In this 
description the $U(1)_R$ symmetry of the $\mathcal{N}=2$ subalgebra of the $\mathcal{N}=6$ manifest 
superconformal symmetry of the theory has weights $(1,1,1,1)$ on $\C^4$, which gives 
a different Hopf action on $\C^4$. In \cite{ABJM} the variables $C_\alpha\equiv(A_1,A_2,B_1^*,B_2^*)$ 
were also used. In this choice of complex structure on $\R^8\cong \C^4$ the $U(1)_M$ and $U(1)_R$ 
weights above are interchanged; in these variables the $SU(4)_R$ symmetry of the theory, 
which acts isometrically on $\mathbb{CP}^3=\{S^7\subset\C^4\}/U(1)_M$, is manifest. However, to be 
uniform with the other examples we shall study, we shall fix the first complex structure on $\R^8\cong\C^4$ above.

In these coordinates $S^7=\{(z_1,z_2,z_3,z_4)\in\C^4\mid |z_1|^2+|z_2|^2+|z_3|^2+|z_4|^2=1\}$, while the 
M-theory Hamiltonian function on $S^7/\Z_k$ is
\bea\label{hMABJM}
h_M &=& \frac{1}{k}\left(|z_1|^2+|z_2|^2-|z_3|^2-|z_4|^2\right)~.
\eea
In the toric geometry language of section \ref{sec:geometry}, we have 
the symplectic coordinates $y_i=\frac{1}{2}|z_i|^2$.
The level sets $h^{-1}_M(c)$ are diffeomorphic to $S^3\times S^3/\Z_k$ for $c\in (-\frac{1}{k},\frac{1}{k})$. 
Perhaps the easiest way to explain this is to note that dividing the levels sets also by $U(1)_M$ 
gives the K\"ahler quotient description of $T^{1,1}$.\footnote{Of course this is directly related to the 
construction of the ABJM theory itself, as the M-theory lift of the theory on $N$ D2-branes at the conifold singularity 
$C(T^{1,1})$, with $k$ units of RR two-form flux through the vanishing $S^2$.} The level sets 
are then a circle bundle over $T^{1,1}\cong S^2\times S^3$, with first Chern class $k\in\Z\cong H^2(S^2\times S^3,\Z)$, 
which means they are diffeomorphic to $S^3\times S^3/\Z_k$.
Notice that these level sets are also described by
\bea
|z_1|^2+|z_2|^2 &=& \frac{1}{2}(1+ck)~, \qquad |z_3|^2+|z_4|^2 \ = \ \frac{1}{2}(1-ck)~.
\eea
When $c\rightarrow \pm \frac{1}{k}$ the $S^3\times S^3/\Z_k$ level sets thus collapse to 
two copies of $S^3/\Z_k$ at $\{z_3=z_4=0\}$ and $\{z_1=z_2=0\}$, respectively. 
Thus the image $h_M(S^7)=\left[-\frac{1}{k},\frac{1}{k}\right]$, with the endpoints 
$\cmax=-\cmin = \frac{1}{k}$
being 
the only two critical points of the Morse-Bott function $h_M$. 

The contact form in these coordinates is
\bea
\ssigma &=& \frac{\ii}{2r^2}\sum_{i=1}^4\left( z_i\diff\bar{z}_i-\bar{z}_i\diff z_i\right)~, \qquad r^2 \ \equiv \ \sum_{i=1}^4 |z_i|^2~.
\eea
Being Einstein, the contact volume of $S^7/\Z_k$ is equal to the Riemannian volume, with 
\bea
\Vol(S^7/\Z_k) &=& \frac{\pi^4}{3k}~.
\eea
Our general formula (\ref{xc}) thus implies that the matrix model variable $x$ should be related to the geometric quantity 
$c$ above via
\be\label{xcABJM}
x \ = \  \frac{(2\pi)^3}{\sqrt{96\, \Vol(S^7/\Z_k)}}\, c \ = \ \pi\sqrt{2k}\, c~.
\ee

The large $N$ saddle point eigenvalue distribution for the ABJM theory was given in \cite{Herzog:2010hf}. 
The eigenvalues for the two gauge groups are related by 
\bea
\lambda^1(x) &=& \bar{\lambda}^2(x) \ = \ x N^{1/2} + \ii y(x)~,
\eea
where
\bea
\rho(x) &=& \frac{\sqrt{k}}{2\pi\sqrt{2}}~, \qquad y(x) \ = \ \frac{\sqrt{k}}{2\sqrt{2}}x~,
\eea
and the eigenvalues are supported on $[\xmin,\xmax]$, where 
$\xmax=-\xmin = \pi \sqrt{2/k}$.
\begin{figure}[ht!]
\centering
\includegraphics[width=0.47\textwidth]{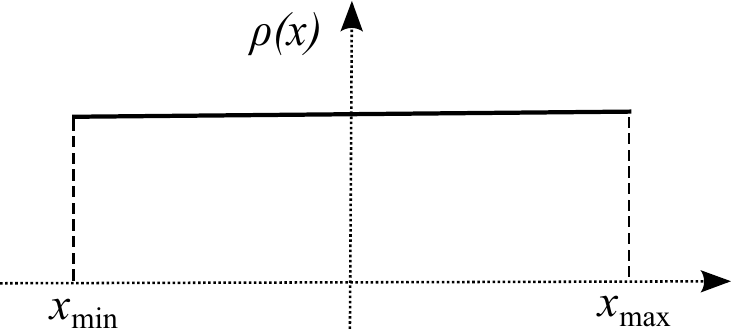}
\caption{Eigenvalue density $\rho(x)$ for the ABJM theory.}\label{rhoABJM}
\end{figure}
This of course agrees with the geometric 
formula (\ref{xcABJM}), and since the density $\rho(x)$ is constant on $(\xmin,\xmax)$ (see Figure \ref{rhoABJM})
its derivative is in particular continuous on this region. It is then automatic that the 
gravity formula (\ref{Wgravity}) agrees with the field theory formula (\ref{Wxmax}) 
for the Wilson loop, giving in both cases
\bea\label{WABJM}
\log \ \langle W \rangle &=& \pi \sqrt{\frac{2}{k}}N^{1/2}~.
\eea

\subsubsection{Mirror theory}\label{sec:mirror}

As discussed in section \ref{sec:Mcircle}, the mirror to the ABJM theory (with $k=1$) arises by 
choosing a different M-theory circle action on $S^7$. The field theory 
\cite{Benini:2009qs} is $\mathcal{N}=8$ $U(N)$ super-Yang-Mills theory 
coupled to two additional fields $q$, $\tilde{q}$ in the fundamental 
and anti-fundamental representation of $U(N)$, respectively. 
The superpotential is 
\bea
\mathcal{W} & =& \mathrm{Tr}\, \left(qX_1\tilde{q} + X_3[X_1,X_2]\right)~,\eea 
where $X_1,X_2,X_3$ are the adjoint chiral fields of the $\mathcal{N}=8$ theory, in 
$\mathcal{N}=2$ language. 

In this case the M-theory circle $U(1)_M$ has weights $(1,-1,0,0)$ on $S^7\subset \C^4$, 
which has a codimension four fixed point set $\FF=S^3=\{z_1=z_2=0\}\subset S^7$. 
It follows that the type IIA internal space is $M_6=S^6$, with a space-filling 
D6-brane wrapping a copy of $S^3\subset S^6$. The field theory described in the previous 
paragraph is then the theory on $N$ D2-branes in flat space ($\mathcal{N}=8$ 
super-Yang-Mills), but coupled to additional massless fields 
$q$, $\tilde{q}$ arising from the lowest excitations of  strings
stretching between the D2-branes and D6-brane.

Although the background geometry is the same as in the previous subsection,
the M-theory Hamiltonian is now\footnote{We could similarly choose to quotient 
by $\Z_k\subset U(1)_M$, which would lead to $k$ D6-branes in the type IIA description. However, here we restricted to $k=1$
in order to compare to the $k=1$ ABJM theory, which is also dual to AdS$_4\times S^7$ (the point being that the 
$\Z_k$ quotients in each case are different). In fact the general $k$ case is $a=k$, $b=0$ of section \ref{sec:flavour}.}
\bea
h_M &=& |z_1|^2 - |z_2|^2~.
\eea
The level surfaces $h_M^{-1}(c)$ are described by 
\bea\label{mirrorlevels}
2|z_1|^2 + |z_3|^2 + |z_4|^2  &=& 1+c~, \qquad 
2|z_2|^2 + |z_3|^2 + |z_4|^2  \ = \ 1-c~,
\eea
so that $c\in [-1,1]$. However, the critical point set of $h_M$ is quite different to that 
for the ABJM model (\ref{hMABJM}). The endpoints $c=+1$, $c=-1$ are 
now the copies of $S^1\subset S^7$ at $\{z_2=z_3=z_4=0\}$ and 
$\{z_1=z_3=z_4=0\}$, respectively. (Compare to the ABJM model, 
where for $k=1$ also $c\in[-1,1]$, but with the endpoints being images of copies of $S^3$, 
rather than $S^1$.) Moreover, there is an \emph{additional} critical point at $c=0$, which then 
intersects the D6-brane locus/fixed point set at $\{z_1=z_2=0\}$. Indeed, 
on $S^7$ we have
\bea
\diff h_M &=& (z_1\diff\bar{z}_1+\bar{z}_1\diff z_1)-(z_2\diff\bar{z}_2+\bar{z}_2\diff z_2)~,\nonumber\\
0 &=& \sum_{i=1}^4 (z_i\diff\bar{z}_i+\bar{z}_i\diff z_i) \qquad \Leftrightarrow \qquad 0 \ = \ \diff r~.
\eea
Thus in addition to the endpoints  $\{z_2=z_3=z_4=0\}$ and 
$\{z_1=z_3=z_4=0\}$, we also have $\diff h_M=0$ at $\{z_1=z_2=0\}=S^3$, which is the fixed point
set of $U(1)_M$ where $h_M=0$. Thus we have the three 
critical points $c_1=\cmin=-1$, $c_2=0$, $c_3=\cmax=1$.

The topology of the level sets $h^{-1}_M(c)$ is the same for $c\in(-1,0)$ and 
$c\in (0,1)$, but with different circles collapsing on each side. 
For $c\in (0,1)$ we may ``solve'' $h_M=c$ as $|z_1|^2 = |z_2|^2+c>0$, and note that consequently 
$z_1\neq 0$ on this locus. From (\ref{mirrorlevels})  it follows that 
$h_M^{-1}(c)\cong S^1_{1}\times S^5$, where $S^1_{1}$ is parametrized by the 
phase of $z_1=|z_1|\ex^{\ii \phi_1}$. On the other hand, for $c\in (-1,0)$ 
instead we solve $h_M=c$ as $|z_2|^2 = |z_1|^2 - c>0$, so that 
$h_M^{-1}(c)\cong S^1_{2}\times S^5$, where $S^1_{2}$ is parametrized by the 
phase of $z_2=|z_2|\ex^{\ii \phi_2}$.

The general formula (\ref{xc}) implies that the matrix model variable $x$ should be related to the geometric quantity 
$c$ again via
\be\label{xcmirror}
x \ = \  \frac{(2\pi)^3}{\sqrt{96\, \Vol(S^7)}}\, c \ = \ \pi\sqrt{2}\, c~,
\ee
which is the same formula as for the ABJM model with $k=1$.
The large $N$ saddle point eigenvalue distribution is in fact a special 
case of the models in section \ref{sec:flavour}, with $a=1$, $b=0$ in the notation of that section, and appears in 
\cite{Jafferis:2011zi}. In this case there is only a single gauge group, 
and one finds the eigenvalue density
\bea
\rho(x) &=& \begin{cases}\frac{1}{2\pi^2}(x-\xmin)~, & \xmin<x<0 \\\frac{1}{2\pi^2}(\xmax-x)~, & 0<x<\xmax \end{cases}~,
\eea
where $\xmax=-\xmin=\pi\sqrt{2}$, thus agreeing with (\ref{xcmirror}).
\begin{figure}[ht!]
\centering
\includegraphics[width=0.47\textwidth]{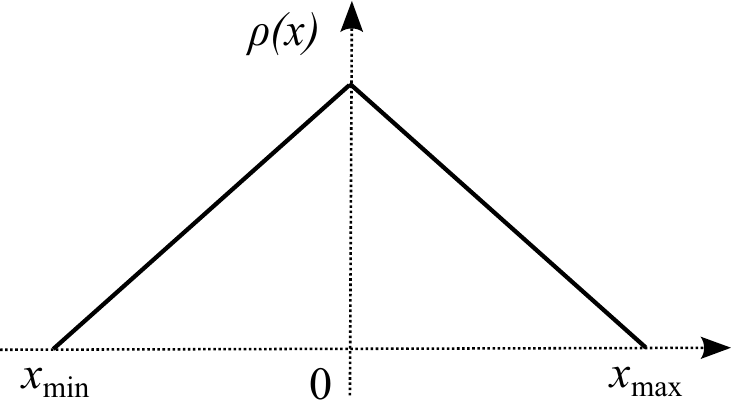}
\caption{Eigenvalue density as a function of $x$. There are three points where $\rho'(x)$ is discontinuous, corresponding to critical points of $h_M$.}
\end{figure} 
Moreover, the derivative of $\rho$ is discontinuous at the endpoints \emph{and} at the point $x=0$, which 
arises as the image of the D6-brane locus under $h_M$. The Wilson loop is again given by 
(\ref{WABJM}), with $k=1$.

\subsection{Dual to $Q^{1,1,1}/\mathbb{Z}_k$}

Our next example is that of the  homogeneous (and toric) Sasaki-Einstein manifold $Q^{1,1,1}/\mathbb{Z}_k$. The manifold $Q^{1,1,1}$ is the total space of an $S^1$ fibration over the product of three copies of $S^2$, {\it  i.e.} $S^1\hookrightarrow Q^{1,1,1}\to S^2\times S^2\times S^2$, which 
describes its structure as a regular Sasaki-Einstein manifold. Even though this manifold is toric, and the geometrical techniques described in section \ref{sec:geometry} can be applied, we will instead take advantage of the fact that the metric is known explicitly on this space. 

The Sasaki-Einstein metric on $Q^{1,1,1}$ can be written as
\be\label{gQ111}
g_{Y_7} \ = \ \frac{1}{16}\left(\dd\psi+\sum_{i=1}^3\cos{\theta_i}\dd\varphi_i\right)^2+\frac{1}{8}\sum_{i=1}^3(\dd\theta_i^2+\sin^2{\theta_i}\dd\varphi_i^2)~,
\ee
where the coordinates $\theta_i\in[0,\pi]$ and $\varphi_i\in[0,2\pi)$ are the usual $S^2$ coordinates, and the coordinate $\psi\in[0,4\pi)$ parametrizes the $S^1$ fibre. The contact form is simply
\be\label{etaQ111}
\ssigma \ = \ \frac{1}{4}\left(\dd\psi+\sum_{i=1}^3\cos{\theta_i}\dd\varphi_i\right)~,
\ee
and for the field theory model below the M-theory circle is generated by $\zeta_M=\frac{1}{k}(\de_{\varphi_1}+\de_{\varphi_2})$. The M-theory Hamiltonian follows straightforwardly and reads
\be
h_M \ = \ \ssigma(\zeta_M) \ = \ \frac{1}{4k}(\cos{\theta_1}+\cos{\theta_2})~.
\ee

The length of a supersymmetric M-theory circle is always given by $2\pi h_M(\hat{p})$, where $\hat{p}\in Y_7$ covers a fixed point $p$ of $\xi$, with $p\in M_6=Y_7/U(1)_M$. 
However, when the Sasaki-Einstein manifold is \emph{regular}, as in the case at hand, we may also describe the 
supersymmetric M-theory circles in terms of the base K\"ahler-Einstein manifold $B_6=Y_7/U(1)_R$, where $U(1)_R$ is generated by the Reeb vector $\xi$. In this point of view, the supersymmetric M-theory circles cover fixed points of $\zeta_M$ on 
$B_6$, which in the case at hand is $B_6=S^2\times S^2\times S^2$ because $\xi=4\partial_\psi$.
These points are located at $\{(\theta_1,\theta_2)\mid \theta_1\in \{0,\pi\},\ \theta_2 \in \{0,\pi\}\}$. Thus one obtains three critical values $c_1=\cmin=-\frac{1}{2k}$, $c_2=0$, $c_3=\cmax=\frac{1}{2k}$. Notice these are $S^2$ loci of critical points, 
parametrized by $(\theta_3,\varphi_3)$.

 Being Einstein, the contact volume of $Q^{1,1,1}/\mathbb{Z}_k$ is equal to the Riemannian volume, with
\be
\Vol(Q^{1,1,1}/\mathbb{Z}_k)\ =\ \frac{\pi^4}{8k}\ ,
\ee
and as usual the $\Z_k$ quotient is along $U(1)_M$ generated by $\zeta_M$.
The general formula \eqref{xc} tells us that the matrix model variable $\xmax=-\xmin$ predicted from the gravity calculation is
\be\label{xcQ111}
\xmax \ = \ \frac{(2\pi)^3}{\sqrt{96\, \Vol(Q^{1,1,1}/\mathbb{Z}_k)}}\, \cmax \ =\ \frac{2\pi}{\sqrt{3k}}\ .
\ee

A dual field theory to $Q^{1,1,1}/\mathbb{Z}_k$ has been proposed in \cite{Jafferis:2009th,Benini:2009qs}. This theory is closely related to the ABJM theory. In addition to the bifundamental fields $A_i$, $B_i$, a pair of field in the (anti-) fundamental representation is added to each gauge group node, and one adds a cubic term to the superpotential
\be
\mathcal{W}_{\mathrm{cubic}} \ = \ \Tr \left(q_1 A_1\tilde q_1+q_2 A_2\tilde q_2\right)~.
\ee
The corresponding matrix model has been worked out in \cite{Cheon:2011vi}, where it was found that the density of the real part of the eigenvalues is
\be
\rho(x) \ = \ \frac{k}{4\pi^2}(2 \xmax-|x|)\quad\mathrm{for}\quad \xmin<x<\xmax~,
\ee
with $\xmax=\frac{2\pi}{\sqrt{3k}}$, thus agreeing with \eqref{xcQ111}.
\begin{figure}[ht!]
\centering
\includegraphics[width=0.47\textwidth]{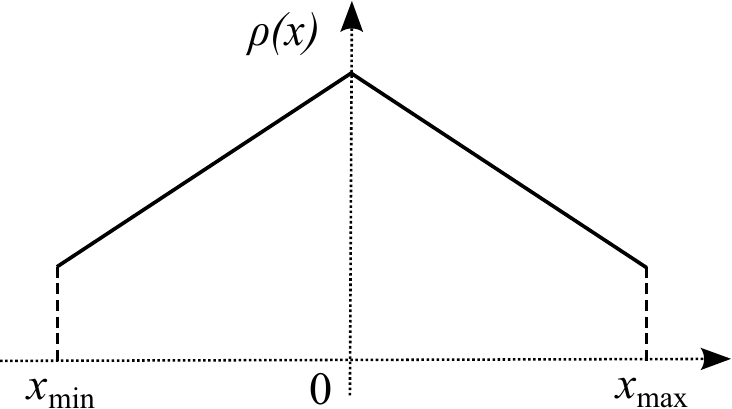}
\caption{Eigenvalue density $\rho(x)$. There are three points where $\rho'(x)$ is discontinuous, associated with supersymmetric M-theory circles.}
\end{figure}
Moreover, the derivative of $\rho$ is discontinuous at the endpoints and at the point $x=0$, as predicted by $c_1, c_2$ and $c_3$ above. The Wilson loop calculated from the field theory then agrees with the gravity computation, and reads
\be
\log \ \langle W \rangle \ = \ \frac{2\pi}{\sqrt{3k}}N^{1/2}\ .
\ee

\subsection{$\mathcal{N}=8$ super-Yang-Mills with flavour}\label{sec:flavour}

In this section we consider a family of $\mathcal{N}=2$ Chern-Simons-matter theories 
that generalize the mirror to the ABJM theory discussed in section \ref{sec:mirror}. 
These were discussed in section 4 of \cite{Jafferis:2011zi}, having been first 
introduced in \cite{Benini:2009qs}. 

One begins 
with $\mathcal{N}=8$ super-Yang-Mills with gauge group $U(N)$, which is the 
theory on $N$ D2-branes in flat space. In $\mathcal{N}=2$ language 
we have three adjoint chiral matter fields $X_1, X_2, X_3$, together with 
the cubic superpotential $\mathrm{Tr}\, X_3[X_1,X_2]$. To this 
we add matter fields in the fundamental and anti-fundamental 
representations, which breaks the supersymmetry generically to $\mathcal{N}=2$. More precisely, 
we add $n_1$ fields $(q_j^{(1)},\tilde{q}^{(1)}_j)$, $n_2$ fields $(q_j^{(2)},\tilde{q}^{(2)}_j)$ 
and $n_3$ fields $(q_j^{(3)},\tilde{q}^{(3)}_j)$, together with the cubic superpotential
\bea\label{Wflavour}
\mathcal{W} &=& \mathrm{Tr}\left[\sum_{j=1}^{n_1}q^{(1)}_jX_1\tilde{q}^{(1)}_j + \sum_{j=1}^{n_2}q^{(2)}_jX_2\tilde{q}^{(2)}_j
+ \sum_{j=1}^{n_3}q^{(3)}_jX_3\tilde{q}^{(3)}_j + X_3[X_1,X_2]\right] ,
\eea
so that the theory in section \ref{sec:mirror} is simply $n_1=1$, $n_2=n_3=0$.
As 
for the mirror to the ABJM theory, the additional $(q,\tilde{q})$ fields will arise in type IIA from 
strings stretching from the D2-branes to D6-brane loci. 

In \cite{Benini:2009qs} it was shown that the (quantum corrected) moduli space 
of vacua of these theories, for $N=1$, may be parametrized by the three 
coordinates $X_1, X_2, X_3$, together with the monopole operators $T$, $\tilde{T}$, 
which satisfy the constraint
\bea\label{modulioneggfl}
T \tilde{T} &=& X_1^{n_1}X_2^{n_2} X_3^{n_3}~.
\eea
This defines a Calabi-Yau cone $C(Y_7)$ as a hypersurface singularity in $\C^5$. 
The M-theory circle is straightforward to identify in this case, since by definition 
the monopole operators $T$, $\tilde{T}$ have charges $\pm1$, respectively, 
under $U(1)_M$, while the $X_i$ are uncharged. Notice this implies that 
the quotient $C(Y_7)/\C^*_M\cong \C^3$ by the \emph{complexified} M-theory circle (defined 
as a GIT quotient) is simply $\C^3$, parametrized by $X_1, X_2, X_3$, which implies that the type IIA description 
involves $N$ D2-branes in flat space. Moreover, $U(1)_M$ fixes $T=\tilde{T}=0$, which defines 
the surface $\{X_1^{n_1}X_2^{n_2}X_3^{n_3}=0\}\subset \C^3$, which becomes 
a D6-brane locus. This then geometrically engineers the gauge theory described above, with superpotential 
(\ref{Wflavour}).

\begin{figure}[ht!]
\centering
\includegraphics[width=0.47\textwidth]{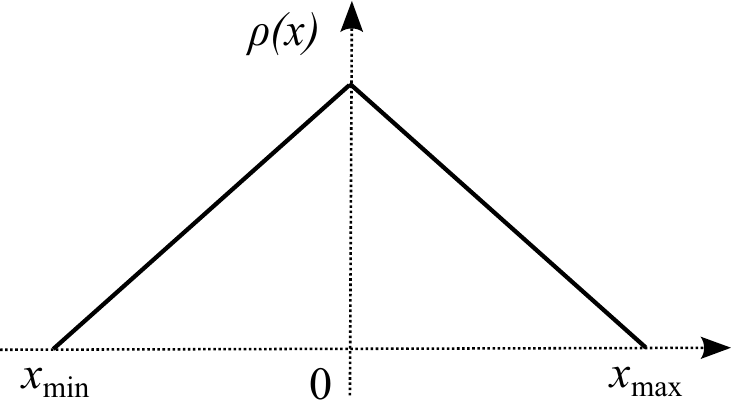}
\caption{Eigenvalue density as a function of $x$. There are three points where $\rho'(x)$ is discontinuous, and we correspondingly expect to find three critical points of $h_M$, with associated supersymmetric circles.}\label{rhoCY3C}
\end{figure}

It is straightforward to analyse the matrix model for this gauge theory, as described in section \ref{sec:localize} 
and carried out in \cite{Jafferis:2011zi}.
The eigenvalue density is given by
\bea
\rho(x) & =& \left\{\begin{array}{lc}
\frac{(\sum_{i=1}^3n_i\Delta_i-2\Delta_m)}{8\pi^2\Delta_1\Delta_2\Delta_3}(x-\xmin)~,&\ \xmin<x<0\\
\frac{(\sum_{i=1}^3n_i\Delta_i+2\Delta_m)}{8\pi^2\Delta_1\Delta_2\Delta_3}(\xmax-x)~,&\ 0<x<\xmax
\end{array}\right.~,
\eea
and the endpoints are 
\bea
x_{\mathrm{max/min}} & =& \pm\sqrt{\frac{8\pi^2\Delta_1\Delta_2\Delta_3(\sum_{i=1}^3n_i\Delta_i\mp 2\Delta_m)}{(\sum_{i=1}^3n_i\Delta_i)(\sum_{i=1}^3n_i\Delta_i\pm 2\Delta_m)}}~.
\eea
Here $\Delta_i=\Delta(X_i)$, $i=1,2,3$, are the R-charges of the fields $X_i$, while $\Delta_m=\Delta(T)=\Delta(\tilde{T})$ 
is the R-charge of the monopole operators. As described in section \ref{sec:localize}, these may be left {\it a priori} 
arbitrary at this point, the only restriction being that the superpotential $\mathcal{W}$ has R-charge $\Delta(\mathcal{W})=2$.
This leads to the constraint $\sum_{i=1}^3\Delta_i=2$.
The shape of $\rho$ as a function of $x$ is shown in Figure \ref{rhoCY3C}.

The superconformal R-charges are determined by maximizing the free energy $F$ as a function of the 
R-charges. This immediately leads to 
$\Delta_m=0$, and then 
\bea
F & = & \frac{2\sqrt{2}\pi \sqrt{\Delta_1\Delta_2\Delta_3\left(\sum_{i=1}^3n_i\Delta_i\right)}}{3}N^{3/2}~,
\eea
which must be further maximized subject to the constraint $\sum_{i=1}^3\Delta_i=2$. 
In practice the formulae are rather too unwieldy for general $n_i$, so 
following \cite{Jafferis:2011zi}
we restrict to the case $n_1=a,n_2=b,n_3=0$. 
In this case the free energy is maximized by
\be
\Delta_1  \ = \ \frac{a-2b+\sqrt{a^2+b^2-ab}}{2(a-b)}~,\;\Delta_2\ =\ \frac{b-2a+\sqrt{a^2+b^2-ab}}{2(b-a)}~,\; \Delta_3 \ = \ \frac{1}{2}\ ,
\ee
and thus
\be\label{xmamxNflacvour}
x_{\mathrm{max/min}}  = \pm 2\pi\sqrt{\frac{\Delta_1\Delta_2}{a\Delta_1+b\Delta_2}}~.
\ee

\begin{figure}[ht!]
\centering
\includegraphics[width=0.50\textwidth]{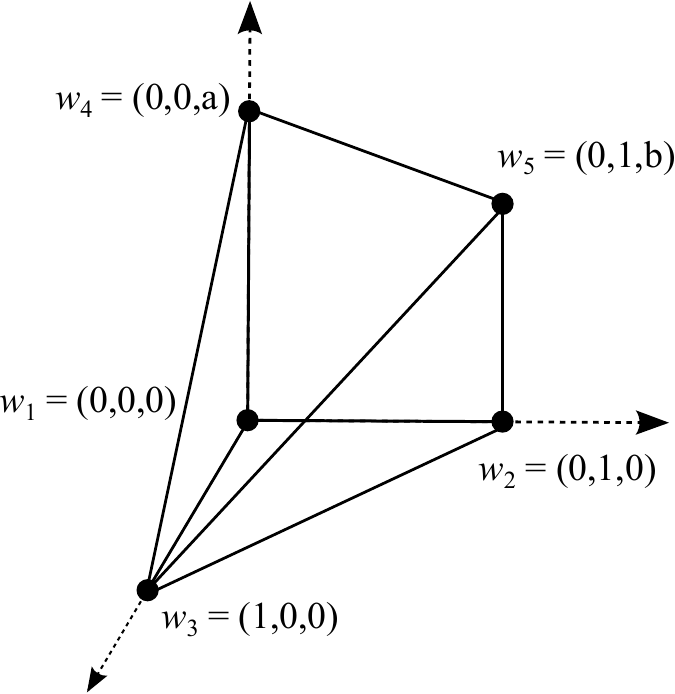}
\caption{Toric diagram corresponding to $C(Y_7)=\{T\tilde T=X_1^aX_2^b\}\times\mathbb{C}$. 
The apex is not an isolated singularity, as one sees from the non-triangular face 
with vertices $(1,2,4,5)$.}\label{toricCY3C}
\end{figure}

The moduli space equation \eqref{modulioneggfl} correspondingly reduces to $T\tilde T=X_1^aX_2^b$. 
The field $X_3$ is then unconstrained, 
and the Calabi-Yau cone  takes the product form $C(Y_7)=\C\times C(Y_5)$, where 
$X_3$ is a coordinate on $\C$ and $C(Y_5)$ is precisely the $Y_5=L^{a,b,a}$ 
toric singularity.\footnote{A discussion of the relation between the hypersurface and toric geometry
descriptions of this three-fold singularity may be found in \cite{Gabella:2010cy}.}
The toric diagram has lattice vectors
\bea\nonumber
&w_1 \ = \ (0,0,0)~, \quad w_2 \ = \ (0,1,0)~, \quad w_3 \ = \ (1,0,0)~,\\
&w_4 \ =\ (0,0,a)~,\quad w_5\ =\ (0,1,b)~,
\eea
and is shown in Figure \ref{toricCY3C}.
Recall that we parametrize the Reeb vector by $\xi=(4,\xi_2,\xi_3,\xi_4)$, and that the four-dimensional outward-pointing vectors to the facets 
are $v_a=(1,w_a)$. With the method described earlier in section \ref{sec:geometry}, the volume of the base $Y_7$ and the Reeb vector can be found and expressed in terms of $\Delta_1$ and $\Delta_2$, and one finds
\bea
\Vol(Y_7) \ = \ \frac{\pi^4}{6}\frac{1}{\Delta_1\Delta_2(a\Delta_1+b\Delta_2)}~,
\eea
and
\bea
\vec\xi & =& \left(4,1,2\Delta_2,a\Delta_1+b\Delta_2\right)~.
\eea

The M-theory circle in this basis is given by $\zeta_M=(0,0,0,-1)$; one can derive this 
by writing the functions $T,\tilde{T}, X_i$ in terms of the toric geometry formalism 
above (see, for example, section 4.3 of \cite{Gabella:2010cy}). Recall also that in this formalism the 
M-theory Hamiltonian function is given by (\ref{Hamtoric}). Thus in this case we have simply $h_M=-2y_4$. 
The critical points of $h_M$ must always lie on the boundary of the polyhedral cone, which are 
coordinate singularities, and thus it is easiest to determine this critical set using the method 
described at the end of section \ref{sec:geometry}.
We denote the face of the toric diagram which has vertices $\{v_a,\ v_b,\ v_c,\ldots\}$  by $(a,b,c,\ldots)$. Equation \eqref{zetaxilin} then has two types of solution:
\bea\nonumber
  h_M \ = \ 0 \qquad \quad \quad \  &\quad \mathrm{on}&\quad(2,3,5),\ (1,3,4),\ (1,2,4,5)~,\\
|h_M| \ = \ \frac{1}{a\Delta_1+b\Delta_2}&\quad \mathrm{on}&\quad (1,2,3),\ (3,4,5)~,
\eea
and correspondingly one has the critical values $h_M=c_i$ given by
\bea
c_3 & =& -c_1 \ = \ \frac{1}{a\Delta_1+b\Delta_2}~,\quad\mathrm{and}\quad c_2 \ = \ 0~.
\eea
Notice here that the face $(1,2,4,5)$ (being non-triangular) corresponds to the $S^1$ locus 
of $L^{a,b,a}$ conical singularities in $Y_7$. 
Using the general formula (\ref{xc}) we then find that these values of $c_i$ precisely match the corresponding positions 
$x_1, x_2, x_3$ at which the derivative of the eigenvalue density $\rho'(x)$ is discontinuous. 
Finally, using \eqref{M2action} the Wilson loop is 
\bea
\log\ \langle W \rangle_{\mathrm{gravity}} &= &2\pi \sqrt{\frac{\Delta_1 \Delta_2}{a \Delta_1+ b\Delta_2}}N^{1/2} \ = \ \xmax N^{1/2} \ = \ \log\
\langle W \rangle_{\mathrm{QFT}}~,
\eea
where we used \eqref{xmamxNflacvour}.

\subsection{$L^{a,2a,a}$ Chern-Simons-quivers}\label{sec:La}

In this section and the next we study two families of examples whose matrix models were first 
analyzed in \cite{Amariti:2012tj}. 

The $\mathcal{N}=2$ field theories begin 
life as low-energy theories on $N$ D2-branes at an $L^{a,b,a}$ Calabi-Yau three-fold singularity. 
This may be simply described as the hypersurface $\{wz=u^av^b\}\subset \C^4$, where 
$(w,z,u,v)$ are the coordinates on $\C^4$.
This geometry also appeared in the previous subsection of course, but there the M-theory Calabi-Yau 
four-fold was a product $\C\times C(L^{a,b,a})$, whereas here instead $C(L^{a,b,a})$ arises 
as the type IIA spacetime. The low-energy theory on the $N$ D2-branes is known from 
\cite{Franco:2005sm, Benvenuti:2005ja, Butti:2005sw}, and is described by a $U(N)^{a+b}$ quiver gauge theory, with
a superpotential $\mathcal{W}$ consisting of both cubic and quartic terms in the 
bifundamental and adjoint chiral matter fields. Without loss of generality we may take $b\geq a$, 
in which case there are $b-a$ adjoint chiral superfields associated to $b-a$ 
of the $a+b$ $U(N)$ gauge group factors, and a total of $2(a+b)$ bifundamental fields. We refer 
the reader to the above references for further details of these quiver gauge theories.

The D2-brane theories become M2-brane theories at a Calabi-Yau four-fold 
by turning on RR flux in the type IIA background, following \cite{Aganagic:2009zk} and in particular the construction in \cite{Ueda:2008hx}.
 Geometrically this fibres 
the M-theory circle over the base $C(L^{a,b,a})$, and in the field theory introduces 
Chern-Simons couplings for the gauge group, described by a vector of Chern-Simons levels 
$\vec k = (k_1,\ldots,k_{a+b})=(k_1,\ldots,k_{b-a}\| k_{b-a+1},\ldots,k_{a+b})$, where the double bar 
separates the copies of $U(N)$ with adjoint fields from those without. This construction is described in more detail in \cite{Amariti:2012tj}.

\begin{figure}[ht!]
\centering
\includegraphics[width=0.54\textwidth]{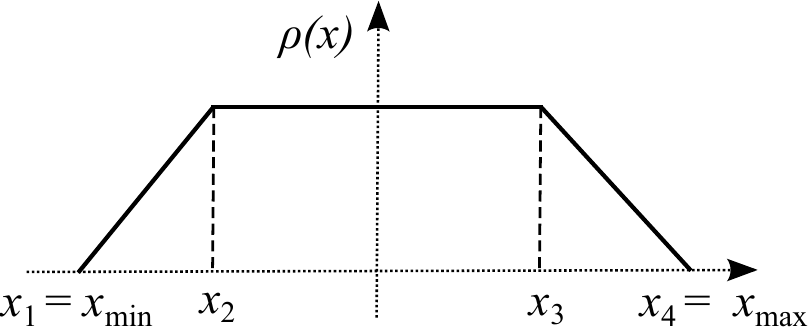}
\caption{Eigenvalue density as a function of $x$. There are 4 points $x_1, x_2, x_3, x_4$ where $\rho'(x)$ is discontinuous, corresponding 
to critical points of $h_M$.}\label{jkhsd}
\end{figure}

Our first class of examples arise from $L^{a,2a,a}$ quiver theories, where the vector of Chern-Simons levels 
is $\vec k = (0,\ldots,0,-2k\| k, k,-k,k,-k,\ldots,k,-k,k)$, with $k\in\Z$. These theories generalize 
the model first studied in \cite{Martelli:2011qj}.
The matrix model 
may be solved using the general large $N$ saddle point method described in section 
\ref{sec:localize}, and one finds \cite{Amariti:2012tj} the eigenvalue density
\bea
\rho(x) & =& \left\{\begin{array}{lc}
\frac{4ak\pi x(1-\Delta)+\mu}{16a\pi^3(1-\Delta)\Delta^2}~,& \ \  -\frac{\mu}{4ak\pi(1-\Delta)}<x<-\frac{\mu}{2ak\pi(2-\Delta)}\\
\frac{\mu}{16 a\pi^3\Delta(2-\Delta)(1-\Delta)}~,& -\frac{\mu}{2ak\pi(2-\Delta)}<x<\frac{\mu}{2ak\pi(2-\Delta)}\\
-\frac{4ak\pi x(1-\Delta)-\mu}{16a\pi^3(1-\Delta)\Delta^2}~,&  \  \ \frac{\mu}{2ak\pi(2-\Delta)}<x<\frac{\mu}{4ak\pi(1-\Delta)}
\end{array}\right.~,
\eea
where we have defined\footnote{The variable $\mu$ arises as a Lagrange multiplier, enforcing that $\rho(x)$ is a density 
satisfying (\ref{density}).}
\bea
\mu & =& 8a\pi^2\sqrt{\frac{k\Delta(2-3\Delta+\Delta^2)^2}{4-3\Delta}}~.
\eea
Here the single R-charge variable $\Delta$ parametrizes the R-charges of all the chiral matter fields, 
as in \cite{Amariti:2012tj}. The eigenvalue density $\rho(x)$ is shown in Figure \ref{jkhsd}.

The free energy, as a function of $\Delta$, is given by
\bea
F & =& \frac{8a\pi }{3}\sqrt{\frac{k\Delta(1-\Delta)^2(2-\Delta)^2}{(4-3\Delta)}}N^{3/2}~.
\eea
One may then maximize $F$ to determine the superconformal $\Delta$, finding the cubic irrational
\bea
\Delta &=& \frac{1}{18}\left[19-\frac{37}{\left(431-18\sqrt{417}\right)^{1/3}}-\left(431-18\sqrt{417}\right)^{1/3}\right]\ \simeq \ 0.319~.\eea
This agrees with the value computed in \cite{Martelli:2011qj}, which was for the particular case $a=1$. 

\begin{figure}[ht!]
\centering
\includegraphics[width=0.55\textwidth]{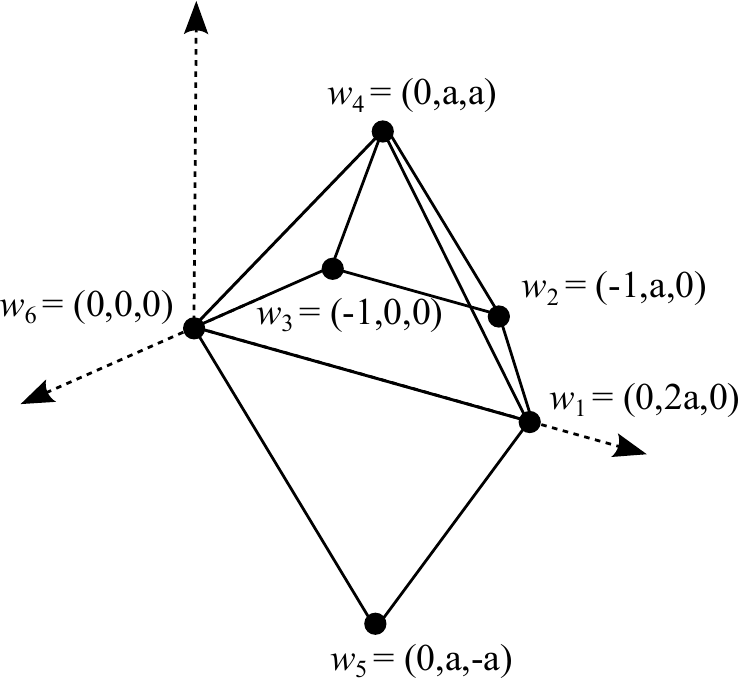}
\caption{Toric diagram of the $L^{a,2a,a}$ Chern-Simons-quiver theories with $k=1$.}\label{oih}
\end{figure}

Turning to the dual geometry, the Calabi-Yau four-fold that arises as the Abelian $N=1$ moduli space of these theories 
has toric data (for $k=1$)
\bea\nonumber
w_1 \ = \ (0,2a,0)~,\quad w_2\ = \ (-1,a,0)~, \quad w_3 \ = \ (-1,0,0)~, \\
w_4 \ = \ (0,a,a)~,\quad w_5 \ = \ (0,a,-a)~, \quad w_6\ = \ (0,0,0)~,
\eea
and with toric diagram shown in Figure \ref{oih}.
The volume of $Y_7$ may be computed as described in section \ref{sec:geometry}, and one obtains
\bea
\Vol(Y_7) & =& \frac{\pi^4(4-3\Delta)}{96 a^2 k\Delta(\Delta-1)^2(\Delta-2)^2}~,
\eea
with corresponding Reeb vector field
\bea
\vec\xi & =& (4,-4\Delta,2a(2-\Delta),0)~.
\eea

The M-theory circle for this field theory is given in this basis by $\zeta_M=(0,0,0,1)$, so that 
$Y_7$ is given by a $\Z_k$ quotient of the geometry appearing in Figure \ref{oih}, 
with $\Z_k\subset U(1)_M$. We may again compute the critical points of the M-theory Hamiltonian 
$h_M=2y_4$ using the method at the end of section \ref{sec:geometry}.
  Equation \eqref{zetaxilin} has solutions associated to the following faces of the toric diagram:
\bea\nonumber
\Big(h_M \ =\ 0 \qquad \qquad &\quad \mathrm{on}&\quad(1,4,5,6)\Big)~,\\\nonumber
|h_M| \ = \ \frac{1}{4 a(1-\Delta)}&\quad \mathrm{on}&\quad (2,3,4)~,\ (2,3,5)~,\\
|h_M| \ = \ \frac{1}{2 a(2-\Delta)}&\quad \mathrm{on}&\quad (1,2,4)~,\ (1,2,5)~,\ (3,4,6)~,\ (3,5,6)~,
\eea
and correspondingly one has the critical values $h_M=c_i$ given by
\bea
c_4 & =& -c_1\ =\ \frac{1}{4ak(1-\Delta)}~, \quad\mathrm{and}\quad c_3 \ = \ -c_2 \ = \ \frac{1}{2ka(2-\Delta)}~.
\eea
Note here that the face $(1,4,5,6)$ describes a \emph{singular} $S^1$ locus in $Y_7$, and thus 
although $h_M=0$ here, formally leading to zero-action M2-branes, the tangent space is singular.
Using the general formula (\ref{xc}) we then find that these values of $c_i$ precisely match the corresponding positions 
$x_1, x_2, x_3, x_4$ at which the derivative of the eigenvalue density $\rho'(x)$ is discontinuous. 
Explicitly, the actions of M2-branes wrapped on the corresponding calibrated $S^1\subset Y_7$ are then
\bea
-S_{\mathrm{M2}}(c_2)&= \ 4\pi(1-\Delta)\sqrt{\frac{\Delta}{k(4-3\Delta)}}N^{1/2}~,\nonumber\\
\log \, \langle W\rangle \ =\ -S_{\mathrm{M2}}(c_4)& = \ 2\pi(2-\Delta)\sqrt{\frac{\Delta}{k(4-3\Delta)}}N^{1/2}~,
\eea
with the latter determining the Wilson loop VEV, and showing that the field theory and gravity computations of it agree.

\subsection{$L^{a,b,a}$ Chern-Simons-quivers}\label{sec:Lab}

\begin{figure}[ht!]
\centering
\includegraphics[width=0.54\textwidth]{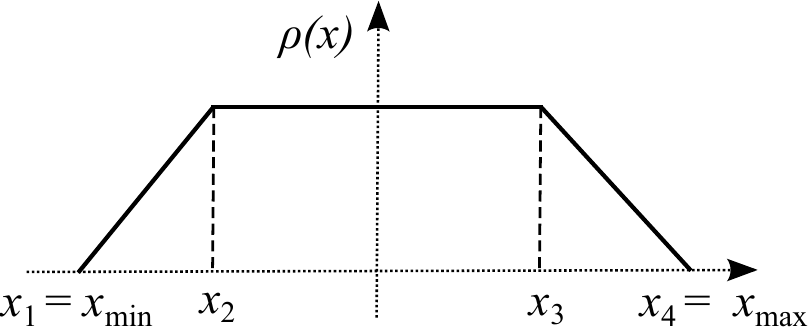}
\caption{Eigenvalue density as a function of $x$. There are again 4 points $x_1, x_2, x_3, x_4$ where $\rho'(x)$ is discontinuous, corresponding 
to critical points of $h_M$.}\label{rho2density}
\end{figure}

Our second family within this class are the $L^{a,b,a}$ Chern-Simons-quiver theories, 
with the vector of Chern-Simons levels now given by $\vec k=(0,\ldots,k,-2k\| k,0,\ldots,0)$.
One finds the eigenvalue density \cite{Amariti:2012tj}
\bea
\rho(x) \  = \  \left\{\begin{array}{lc}
\frac{4k\pi x(1-\Delta)+\mu}{16\pi^3(1-\Delta)\Delta((b-2)(1-\Delta)+a\Delta)}~, & \ \ \ \ \  \ \ \ -\frac{\mu}{4k\pi(1-\Delta)}<x<-\frac{\mu}{2k\pi(b(1-\Delta)+a\Delta)}\\
\frac{\mu}{16\pi^3(1-\Delta)\Delta(b(1-\Delta)+a\Delta)}~, &   -\frac{\mu}{2k\pi(b(1-\Delta)+a\Delta)}<x<\frac{\mu}{2k\pi(b(1-\Delta)+a\Delta)}\\
-\frac{4k\pi x(1-\Delta)-\mu}{16\pi^3(1-\Delta)\Delta((b-2)(1-\Delta)+a\Delta)}~,& \frac{\mu}{2k\pi(b(1-\Delta)+a\Delta)}<x<\frac{\mu}{4k\pi(1-\Delta)}
\end{array}\right.
\eea
where we have defined
\bea
\mu & =& 8\pi^2\sqrt{\frac{k\Delta(1-\Delta)^2(b(1-\Delta)+a\Delta)^2}{(b-2)(1-\Delta)+a\Delta}}~.
\eea
Again, the R-charge variable $\Delta$ parametrizes the R-charges of all the chiral matter fields, 
as detailed in \cite{Amariti:2012tj}. The eigenvalue density $\rho(x)$ is shown in Figure \ref{rho2density}.

The free energy, as a function of $\Delta$, is given by
\bea
F & =& \frac{8\pi}{3}\sqrt{\frac{k(1-\Delta)^2\Delta(b(1-\Delta)+a\Delta)^2}{(b+2)(1-\Delta)+a\Delta}}N^{3/2}~.
\eea
One may then maximize $F$ to find an expression (not presented) for the superconformal $\Delta$ that depends on $a$ and $b$.

\begin{figure}[ht!]
\centering
\includegraphics[width=0.8\textwidth]{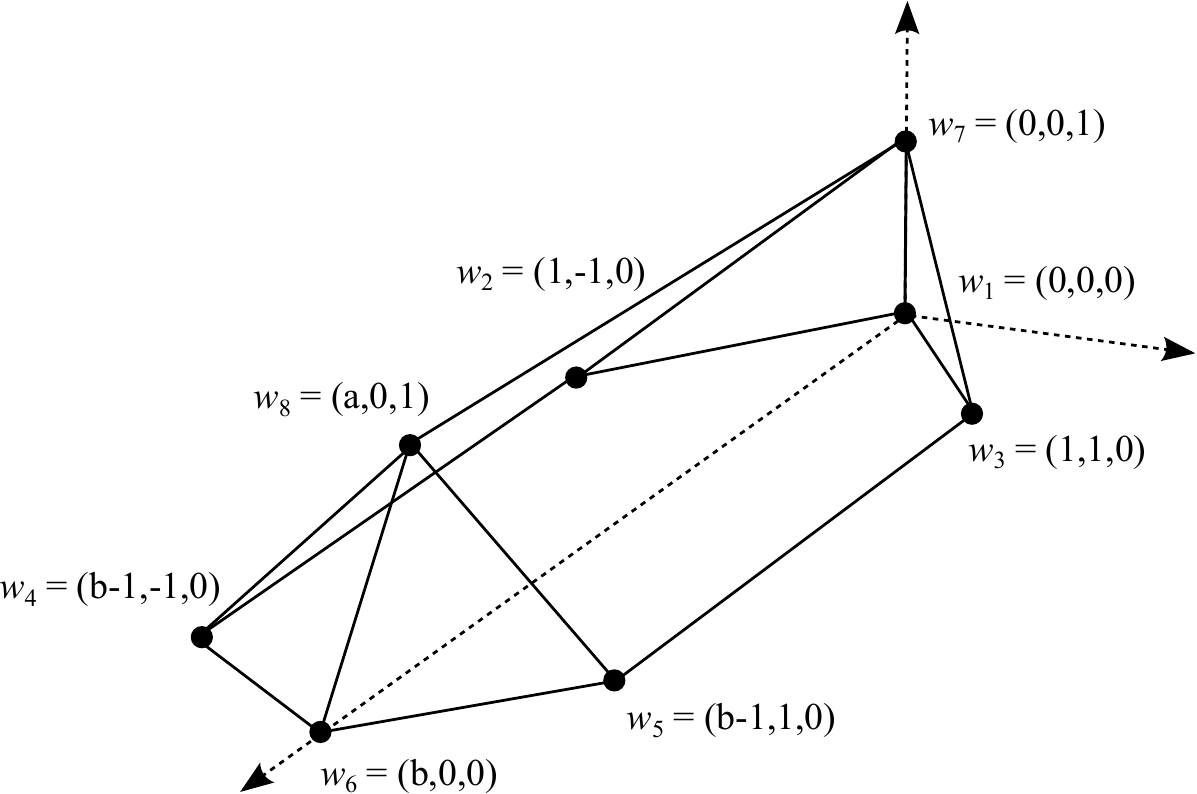}
\caption{Toric diagram of the $L^{a,b,a}$ Chern-Simons-quiver theories with $k=1$.}\label{oih3}
\end{figure}

The corresponding Calabi-Yau four-fold that arises as the Abelian $N=1$ moduli space of these theories 
has toric data (for $k=1$)
\bea\nonumber
w_1 \ =\ (0,0,0)~, \; w_2 \ = \ (1,-1,0)~, \; w_3 \ =\ (1,1,0)~, \; w_4 \ = \ (b-1,-1,0)~,\\
w_5\ =\ (b-1,1,0)~, \; w_6\ =\ (b,0,0)~, \; w_7 \ =\ (0,0,1)~, \; w_8\ =\ (a,0,1)~,
\eea
and with toric diagram shown in Figure \ref{oih3}.
The volume of $Y_7$ may be computed as described in section \ref{sec:geometry}, and one obtains
\bea
\Vol(Y_7) & =& \frac{\pi^4((b+2)(1-\Delta)+a \Delta)}{96k\Delta (1-\Delta)^2(b(1-\Delta)+a\Delta)^2}~,
\eea
with corresponding Reeb vector
\bea
\vec\xi & =& (4,2(b(1-\Delta)+ a\Delta),0,4\Delta)~.
\eea

The M-theory circle for this field theory is given in this basis by $\zeta_M=(0,0,1,0)$, so that again
$Y_7$ is given by a $\Z_k$ quotient of the geometry appearing in Figure \ref{oih3}, 
with $\Z_k\subset U(1)_M$. The M-theory Hamiltonian is 
$h_M=2y_3$, and its critical points may be computed from 
equation \eqref{zetaxilin}, which has solutions on the following faces of the toric diagram:
\bea\nonumber
\Big(h_M\ =\ 0 \  \quad \qquad \qquad \qquad &\quad \mathrm{on}&\quad(1,2,3,4,5,6)~,\Big)\\
|h_M|\ =\ \frac{1}{4 (1-\Delta)} \ \   \quad \qquad &\quad \mathrm{on}&\quad (2,4,7,8)~,\ (3,5,7,8)~,\\\nonumber
|h_M|\ =\ \frac{1}{2 (b(1-\Delta)+a\Delta)} &\quad \mathrm{on}&\quad (1,2,7)~,\ (1,3,7)~,\ (4,6,8)~,\ (5,6,8)~,
\eea
and correspondingly one has critical values $h_M=c_i$ given by 
\bea
c_4 & =& -c_1 \ =\ \frac{1}{4k(1-\Delta)}~, \quad\mathrm{and}\quad c_3 \ =\ -c_2 \ =\ \frac{1}{2k(b(1-\Delta)+a\Delta)}~.
\eea
Using the general formula (\ref{xc}) we then find that these values of $c_i$ precisely match the corresponding positions 
$x_1, x_2, x_3, x_4$ at which the derivative of the eigenvalue density $\rho'(x)$ is discontinuous. 
Explicitly, the actions of M2-branes wrapped on the corresponding calibrated $S^1\subset Y_7$ are
\bea
-S_{\mathrm{M2}}(c_2 ) \ = &  4\pi(1-\Delta)\sqrt{\frac{\Delta}{k((2+b)(1-\Delta)+a\Delta)}}N^{1/2}~,\nonumber\\
\log \, \langle W \rangle \ =\ -S_{\mathrm{M2}}(c_4) \ = & 2\pi(b(1-\Delta+a\Delta))\sqrt{\frac{\Delta}{k((2+b)(1-\Delta)+a\Delta)}}N^{1/2}~,
\eea
with the latter determining the Wilson loop VEV, and showing that the field theory and gravity computations of it agree.

\subsection{Duals to non-Einstein solutions with flux}

In this final class of examples we examine a family of $\mathcal{N}=2$ superconformal theories  
which are dual to warped \emph{non-Einstein} solutions, with non-zero internal $F_4$ flux in 
(\ref{11d}). As we described in section \ref{sec:gravity}, solutions with non-zero M2-brane charge $N$ 
still have a contact structure, and our formalism then applies.

Following \cite{Jafferis:2011zi}, we begin with the $\mathcal{N}=8$ super-Yang-Mills theories with 
flavour in section \ref{sec:flavour}, for which the dual Calabi-Yau four-fold is 
$C(Y_7)=\C\times C(L^{a,b,a})$. The field theories have a free chiral field $X_3$, of scaling dimension 
$\Delta(X_3)=\Delta_3=\frac{1}{2}$,
and we consider perturbing the theory by adding the deformation 
$\lambda X_3^{p}$ to the superpotential. In three dimensions this is a relevant deformation for 
$p=2$ and $p=3$. The gravity dual to the resulting infrared  (IR) fixed point of the massive $p=2$ deformation is the 
Corrado-Pilch-Waner solution of \cite{Corrado:2001nv}, while the supergravity dual to the cubic $p=3$ deformation was 
found only recently (independently) in \cite{Gabella:2012rc, Halmagyi:2012ic}. In both cases these
are warped AdS$_4\times Y_7$ solutions with flux, where crucially $Y_7$ has 
the \emph{same} topology as before the deformation, and the toric symplectic structure 
of the cone $C(Y_7)$ is also the same. However, the Reeb vector field (and hence also contact structure)
is different. 

This deformation was studied in the matrix model in \cite{Jafferis:2011zi}, where they found the universal behaviour
\bea\label{FIR}
\frac{F^{\mathrm{IR}}}{F^{\mathrm{UV}}} &=& \frac{16(p-1)^{3/2}}{3\sqrt{3}p^2}~,
\eea
relating the free energies $F$ of the IR and UV theories. One correspondingly finds that the scaling 
dimensions $\Delta_i=\Delta(X_i)$ are related as
\bea\label{RIR}
\frac{\Delta_1^{\mathrm{IR}}}{\Delta_1^{\mathrm{UV}}} &=& \frac{4(p-1)}{3p} \ =\ 
\frac{\Delta_2^{\mathrm{IR}}}{\Delta_2^{\mathrm{UV}}}~, \qquad \Delta^{\mathrm{IR}}_3 \ = \ \frac{2}{p}~,
\eea
and that the eigenvalue distribution takes the same form as before, with $\xmax=-\xmin$, but now with the endpoints rescaled as
\bea\label{xIR}
\frac{x^{\mathrm{IR}}_{\mathrm{max}}}{x^{\mathrm{UV}}_{\mathrm{max}}} &=& \frac{4\sqrt{p-1}}{\sqrt{3}p}~.
\eea

As explained in general in \cite{Gabella:2012rc}, using (\ref{free}) the field theory result (\ref{FIR}) matches with the supergravity result relating the
(contact) volumes of the IR and UV solutions: 
\bea
\frac{\Vol_\ssigma(Y_7^{\mathrm{IR}})}{\Vol(Y_7^{\mathrm{UV}})} &=& \frac{27p^4}{256 (p-1)^3} \ \equiv \ V(p)~.
\eea
This result was derived in \cite{Gabella:2012rc} using the contact geometry of the IR solutions. 
It is crucial that here the volume is the contact volume $\Vol_\ssigma$ and \emph{not} the Riemannian volume (which is different). 
The Reeb vector field of the UV Calabi-Yau geometry $\C\times C(Y_5)$ may be written as
\bea
\xi^{\mathrm{UV}} &=& \partial_{\phi_3} + \partial_\varphi~,
\eea
where $\partial_{\phi_3}$ rotates the free field $X_3$ with charge 1 (so $X_3=|X_3|\ex^{\ii \phi_3}$), while 
$\partial_\varphi$ denotes the Reeb vector field of the Calabi-Yau three-fold cone $C(Y_5)$. As shown in 
\cite{Gabella:2012rc}, the IR Reeb vector field for the solution with flux is then
\bea
\xi^{\mathrm{IR}} &=& \frac{4}{p}\partial_{\phi_3} + \frac{4(p-1)}{3p}\partial_\varphi~.
\eea
In fact these formulae are directly related to the rescalings of the R-charges in (\ref{RIR}).
Using this simple rescaling one can check that the M2-brane actions are rescaled as
\bea
\frac{S_{\mathrm{M2}}^{\mathrm{IR}}}{S_{\mathrm{M2}}^{\mathrm{UV}}} &=& \frac{\frac{3p}{4(p-1)}}{\sqrt{V(p)}} \ = \  \frac{4\sqrt{p-1}}{\sqrt{3}p}~,
\eea
thus matching the field theory result (\ref{xIR}). It of course follows immediately that the 
field theory and gravity results for the Wilson loop VEV agree. What is remarkable here 
is that we are able to compute this so straightforwardly, even though the 
$p=3$ supergravity solution is known only numerically in \cite{Gabella:2012rc, Halmagyi:2012ic}. 
Of course, it is precisely because we know the \emph{contact structure} explicitly that the computation is straightforward, and this is all that is required
to compute these BPS quantities.

\section{Discussion}\label{sec:discussion}

In this paper we have shown that the large $N$ field theory and gravity computations of the BPS Wilson loop agree 
in a large class of three-dimensional $\mathcal{N}=2$ superconformal field theories with 
AdS$_4\times Y_7$ gravity duals. In fact really this matching is a corollary of the fact that 
the image of the M-theory Hamiltonian $h_M(Y_7)=[\cmin,\cmax]$ is equal to the 
support $[\xmin,\xmax]$ of the real part of the saddle point eigenvalue distribution in the large $N$ matrix model,  
with the proportionality factor between the variables $x$ and $c$ given by (\ref{xc}). Moreover, the critical points of $h_M$, 
which give the loci of supersymmetric M2-branes wrapping the M-theory circle, 
always map under $h_M$ to the points at which $\rho'(x)$ is discontinuous in the matrix model. The fact that the eigenvalue density changes behaviour every time a critical point $x_i$ is crossed is explained by \eqref{rhohm} which relates $\rho(x)$ to the volume of a subspace of $h^{-1}_M(c)$ whose topology changes at the critical points $c_i$.

\subsection*{Acknowledgments}
\noindent
 D.~F. is supported by the Berrow Foundation and J.~F.~S. by the Royal Society. 
We would like to thank Chris Herzog for pointing out the results of reference \cite{Gulotta:2011si, Gulotta:2011aa}.


\begin{thebibliography}{99}

\bibitem{ABJM}
O.~Aharony, O.~Bergman, D.~L.~Jafferis and J.~Maldacena,
  ``${\cal N}=6$ superconformal Chern-Simons-matter theories, M2-branes and their
  gravity duals,''
  JHEP {\bf 0810}, 091 (2008)
  [arXiv:0806.1218 [hep-th]].




\bibitem{Benna:2008zy} 
  M.~Benna, I.~Klebanov, T.~Klose and M.~Smedback,
  ``Superconformal Chern-Simons Theories and AdS(4)/CFT(3) Correspondence,''
  JHEP {\bf 0809}, 072 (2008)
  [arXiv:0806.1519 [hep-th]].

\bibitem{Martelli:2008si} 
  D.~Martelli and J.~Sparks,
  ``Moduli spaces of Chern-Simons quiver gauge theories and AdS(4)/CFT(3),''
  Phys.\ Rev.\ D {\bf 78}, 126005 (2008)
  [arXiv:0808.0912 [hep-th]].

\bibitem{Jafferis:2008qz} 
  D.~L.~Jafferis and A.~Tomasiello,
  ``A Simple class of N=3 gauge/gravity duals,''
  JHEP {\bf 0810}, 101 (2008)
  [arXiv:0808.0864 [hep-th]].

\bibitem{Hanany:2008cd} 
  A.~Hanany and A.~Zaffaroni,
  ``Tilings, Chern-Simons Theories and M2 Branes,''
  JHEP {\bf 0810}, 111 (2008)
  [arXiv:0808.1244 [hep-th]].

\bibitem{Ueda:2008hx} 
  K.~Ueda and M.~Yamazaki,
  ``Toric Calabi-Yau four-folds dual to Chern-Simons-matter theories,''
  JHEP {\bf 0812}, 045 (2008)
  [arXiv:0808.3768 [hep-th]].

\bibitem{Hanany:2008fj} 
  A.~Hanany, D.~Vegh and A.~Zaffaroni,
  ``Brane Tilings and M2 Branes,''
  JHEP {\bf 0903}, 012 (2009)
  [arXiv:0809.1440 [hep-th]].

\bibitem{Franco:2008um} 
  S.~Franco, A.~Hanany, J.~Park and D.~Rodriguez-Gomez,
  ``Towards M2-brane Theories for Generic Toric Singularities,''
  JHEP {\bf 0812}, 110 (2008)
  [arXiv:0809.3237 [hep-th]].

\bibitem{Hanany:2008gx} 
  A.~Hanany and Y.~-H.~He,
  ``M2-Branes and Quiver Chern-Simons: A Taxonomic Study,''
  arXiv:0811.4044 [hep-th].

\bibitem{Franco:2009sp} 
  S.~Franco, I.~R.~Klebanov and D.~Rodriguez-Gomez,
  ``M2-branes on Orbifolds of the Cone over Q**1,1,1,''
  JHEP {\bf 0908}, 033 (2009)
  [arXiv:0903.3231 [hep-th]].

\bibitem{Jafferis:2009th}
  D.~L.~Jafferis,
  ``Quantum corrections to ${\cal N}=2$ Chern-Simons theories with flavor and their AdS4
  duals,''
  arXiv:0911.4324 [hep-th].

\bibitem{Aganagic:2009zk} 
  M.~Aganagic,
  ``A Stringy Origin of M2 Brane Chern-Simons Theories,''
  Nucl.\ Phys.\ B {\bf 835}, 1 (2010)
  [arXiv:0905.3415 [hep-th]].

\bibitem{Benini:2009qs}
  F.~Benini, C.~Closset and S.~Cremonesi,
  ``Chiral flavors and M2-branes at toric CY4 singularities,''
  JHEP {\bf 1002}, 036 (2010)
  [arXiv:0911.4127 [hep-th]].

\bibitem{Amariti:2009rb} 
  A.~Amariti, D.~Forcella, L.~Girardello and A.~Mariotti,
  ``3D Seiberg-like Dualities and M2 Branes,''
  JHEP {\bf 1005}, 025 (2010)
  [arXiv:0903.3222 [hep-th]].

\bibitem{Davey:2009sr} 
  J.~Davey, A.~Hanany, N.~Mekareeya and G.~Torri,
  ``Phases of M2-brane Theories,''
  JHEP {\bf 0906}, 025 (2009)
  [arXiv:0903.3234 [hep-th]].

\bibitem{Hanany:2009vx} 
  A.~Hanany and Y.~-H.~He,
  ``Chern-Simons: Fano and Calabi-Yau,''
  Adv.\ High Energy Phys.\  {\bf 2011}, 204576 (2011)
  [arXiv:0904.1847 [hep-th]].

\bibitem{Davey:2009qx} 
  J.~Davey, A.~Hanany, N.~Mekareeya and G.~Torri,
  ``Higgsing M2-brane Theories,''
  JHEP {\bf 0911}, 028 (2009)
  [arXiv:0908.4033 [hep-th]].

\bibitem{Closset:2012ep} 
  C.~Closset and S.~Cremonesi,
  ``Toric Fano varieties and Chern-Simons quivers,''
  JHEP {\bf 1205}, 060 (2012)
  [arXiv:1201.2431 [hep-th]].




\bibitem{Lunin:2005jy} 
  O.~Lunin and J.~M.~Maldacena,
  ``Deforming field theories with U(1) x U(1) global symmetry and their gravity duals,''
  JHEP {\bf 0505}, 033 (2005)
  [hep-th/0502086].

\bibitem{Corrado:2001nv} 
  R.~Corrado, K.~Pilch and N.~P.~Warner,
  ``An N=2 supersymmetric membrane flow,''
  Nucl.\ Phys.\ B {\bf 629}, 74 (2002)
  [hep-th/0107220].

\bibitem{Klebanov:2008vq} 
  I.~Klebanov, T.~Klose and A.~Murugan,
  ``AdS(4)/CFT(3) Squashed, Stretched and Warped,''
  JHEP {\bf 0903}, 140 (2009)
  [arXiv:0809.3773 [hep-th]].

\bibitem{Ahn:2009bq} 
  C.~Ahn and K.~Woo,
  ``The Gauge Dual of A Warped Product of AdS(4) and A Squashed and Stretched Seven-Manifold,''
  Class.\ Quant.\ Grav.\  {\bf 27}, 035009 (2010)
  [arXiv:0908.2546 [hep-th]].

\bibitem{Ahn:2009sk} 
  C.~Ahn,
  ``The Eleven-Dimensional Uplift of Four-Dimensional Supersymmetric RG Flow,''
  J.\ Geom.\ Phys.\  {\bf 62}, 1480 (2012)
  [arXiv:0910.3533 [hep-th]].

\bibitem{Gabella:2011sg} 
  M.~Gabella, D.~Martelli, A.~Passias and J.~Sparks,
  ``The free energy of ${\cal N}=2$ supersymmetric AdS$_4$ solutions of M-theory,''
  JHEP {\bf 1110}, 039 (2011)
  [arXiv: 1107.5035 [hep-th]].

\bibitem{Gabella:2012rc} 
  M.~Gabella, D.~Martelli, A.~Passias and J.~Sparks,
  ``${\cal N}=2$ supersymmetric AdS$_4$ solutions of M-theory,''
  arXiv:1207.3082 [hep-th].




  
\bibitem{Kapustin:2009kz}
  A.~Kapustin, B.~Willett and I.~Yaakov,
  ``Exact Results for Wilson Loops in Superconformal Chern-Simons Theories with Matter,''
  JHEP {\bf 1003}, 089 (2010)
  [arXiv:0909.4559 [hep-th]].

\bibitem{Jafferis:2010un}
  D.~L.~Jafferis,
  ``The Exact Superconformal R-Symmetry Extremizes $Z$,'' 
  arXiv: 1012.3210 [hep-th].

\bibitem{Hama:2010av}
  N.~Hama, K.~Hosomichi and S.~Lee,
  ``Notes on SUSY Gauge Theories on Three-Sphere,''
  arXiv:1012.3512 [hep-th].

\bibitem{Pestun:2007rz} 
  V.~Pestun,
  Commun.\ Math.\ Phys.\  {\bf 313}, 71 (2012)
  [arXiv:0712.2824 [hep-th]].






\bibitem{Herzog:2010hf}
  C.~P.~Herzog, I.~R.~Klebanov, S.~S.~Pufu and T.~Tesileanu,
  ``Multi-Matrix Models and Tri-Sasaki Einstein Spaces,''
  arXiv:1011.5487 [hep-th].

\bibitem{Martelli:2011qj} 
  D.~Martelli and J.~Sparks,
  ``The large N limit of quiver matrix models and Sasaki-Einstein manifolds,''
  Phys.\ Rev.\ D {\bf 84}, 046008 (2011)
  [arXiv:1102.5289 [hep-th]].

\bibitem{Cheon:2011vi} 
  S.~Cheon, H.~Kim and N.~Kim,
  ``Calculating the partition function of N=2 Gauge theories on $S^3$ and AdS/CFT correspondence,''
  JHEP {\bf 1105}, 134 (2011)
  [arXiv:1102.5565 [hep-th]].

\bibitem{Jafferis:2011zi} 
  D.~L.~Jafferis, I.~R.~Klebanov, S.~S.~Pufu and B.~R.~Safdi,
  ``Towards the F-Theorem: N=2 Field Theories on the Three-Sphere,''
  JHEP {\bf 1106}, 102 (2011)
  [arXiv:1103.1181 [hep-th]].

\bibitem{Gulotta:2011vp} 
  D.~R.~Gulotta, J.~P.~Ang and C.~P.~Herzog,
  ``Matrix Models for Supersymmetric Chern-Simons Theories with an ADE Classification,''
  JHEP {\bf 1201}, 132 (2012)
  [arXiv:1111.1744 [hep-th]].

\bibitem{Amariti:2011jp} 
  A.~Amariti and M.~Siani,
  ``Z Extremization in Chiral-Like Chern Simons Theories,''
  JHEP {\bf 1206}, 171 (2012)
  [arXiv:1109.4152 [hep-th]].

\bibitem{Amariti:2011uw} 
  A.~Amariti, C.~Klare and M.~Siani,
  ``The Large N Limit of Toric Chern-Simons Matter Theories and Their Duals,''
  JHEP {\bf 1210}, 019 (2012)
  [arXiv:1111.1723 [hep-th]].

\bibitem{Amariti:2012tj} 
  A.~Amariti and S.~Franco,
  ``Free Energy vs Sasaki-Einstein Volume for Infinite Families of M2-Brane Theories,''
  JHEP {\bf 1209}, 034 (2012)
  [arXiv:1204.6040 [hep-th]].


 



\bibitem{Drukker:2009hy} 
  N.~Drukker and D.~Trancanelli,
  ``A Supermatrix model for N=6 super Chern-Simons-matter theory,''
  JHEP {\bf 1002}, 058 (2010)
  [arXiv:0912.3006 [hep-th]].

\bibitem{Marino:2009jd}
  M.~Marino and P.~Putrov,
  ``Exact Results in ABJM Theory from Topological Strings,''
  JHEP {\bf 1006}, 011 (2010)
  [arXiv:0912.3074 [hep-th]].
  
\bibitem{Drukker:2010nc}
  N.~Drukker, M.~Marino and P.~Putrov,
  ``From weak to strong coupling in ABJM theory,''
  arXiv:1007.3837 [hep-th].
  
\bibitem{Suyama:2009pd}
  T.~Suyama,
  ``On Large N Solution of ABJM Theory,''
  Nucl.\ Phys.\  B {\bf 834}, 50 (2010)
  [arXiv:0912.1084 [hep-th]].



\bibitem{Maldacena:1998im} 
  J.~M.~Maldacena,
  ``Wilson loops in large N field theories,''
  Phys.\ Rev.\ Lett.\  {\bf 80}, 4859 (1998)
  [hep-th/9803002].

\bibitem{Assel:2012nf} 
  B.~Assel, J.~Estes, M.~Yamazaki and ,
  ``Wilson Loops in 5d N=1 SCFTs and AdS/CFT,''
  arXiv:1212.1202 [hep-th].

\bibitem{Passias:2012vp} 
  A.~Passias,
  ``A note on supersymmetric AdS$_6$ solutions of massive type IIA supergravity,''
  JHEP {\bf 1301}, 113 (2013)
  [arXiv:1209.3267 [hep-th]].


\bibitem{Gabella:2010cy}
  M.~Gabella and J.~Sparks,
  ``Generalized Geometry in AdS/CFT and Volume Minimization,''
  arXiv:1011.4296 [hep-th].

\bibitem{Gabella:2009ni} 
  M.~Gabella, J.~P.~Gauntlett, E.~Palti, J.~Sparks and D.~Waldram,
  ``The Central charge of supersymmetric AdS(5) solutions of type IIB supergravity,''
  Phys.\ Rev.\ Lett.\  {\bf 103}, 051601 (2009)
  [arXiv:0906.3686 [hep-th]].

\bibitem{Gabella:2009wu} 
  M.~Gabella, J.~P.~Gauntlett, E.~Palti, J.~Sparks and D.~Waldram,
  ``AdS(5) Solutions of Type IIB Supergravity and Generalized Complex Geometry,''
  Commun.\ Math.\ Phys.\  {\bf 299}, 365 (2010)
  [arXiv:0906.4109 [hep-th]].

\bibitem{Martelli:2005tp}
  D.~Martelli, J.~Sparks and S.~T.~Yau,
  ``The geometric dual of a-maximisation for toric Sasaki-Einstein manifolds,''
  Commun.\ Math.\ Phys.\  {\bf 268}, 39 (2006)
  [arXiv:hep-th/0503183].

\bibitem{Martelli:2006yb}
  D.~Martelli, J.~Sparks and S.~T.~Yau,
  ``Sasaki-Einstein manifolds and volume minimisation,''
  Commun.\ Math.\ Phys.\  {\bf 280}, 611 (2008)
  [arXiv:hep-th/0603021].

\bibitem{Witten:1991zz} 
  E.~Witten,
  ``Mirror manifolds and topological field theory,''
  In *Yau, S.T. (ed.): Mirror symmetry I* 121-160
  [hep-th/9112056].

\bibitem{Klemm:2012ii} 
  A.~Klemm, M.~Marino, M.~Schiereck and M.~Soroush,
  ``ABJM Wilson loops in the Fermi gas approach,''
  arXiv:1207.0611 [hep-th].
 
\bibitem{Closset:2012vg} 
  C.~Closset, T.~T.~Dumitrescu, G.~Festuccia, Z.~Komargodski and N.~Seiberg,
  ``Contact Terms, Unitarity, and F-Maximization in Three-Dimensional Superconformal Theories,''
  JHEP {\bf 1210}, 053 (2012)
  [arXiv:1205.4142 [hep-th]].

\bibitem{Becker21995}
K.~Becker, M.~Becker ans A.~Strominger,
  ``Fivebranes, membranes and non-perturbative string theory,''
  Nucl.\ Phys.\ B {\bf 456}, 1 (1995)
  [arXiv:hep-th/9507158].

\bibitem{lerman} E.~Lerman, ``Contact toric manifolds,'' J. Symplectic Geom. {\bf 1} (2003), no. 4, 785-828, 
[arXiv:math.SG /0107201].

\bibitem{Futaki:2006cc} 
  A.~Futaki, H.~Ono and G.~Wang,
  ``Transverse Kahler geometry of Sasaki manifolds and toric Sasaki-Einstein manifolds,''
  J.\ Diff.\ Geom.\  {\bf 83}, 585 (2009)
  [math/0607586 [math-dg]].

\bibitem{Benishti:2009ky} 
  N.~Benishti, Y.~-H.~He and J.~Sparks,
  ``(Un)Higgsing the M2-brane,''
  JHEP {\bf 1001}, 067 (2010)
  [arXiv:0909.4557 [hep-th]].

\bibitem{Franco:2005sm} 
  S.~Franco, A.~Hanany, D.~Martelli, J.~Sparks, D.~Vegh and B.~Wecht,
  ``Gauge theories from toric geometry and brane tilings,''
  JHEP {\bf 0601}, 128 (2006)
  [hep-th/0505211].

\bibitem{Benvenuti:2005ja} 
  S.~Benvenuti and M.~Kruczenski,
  ``From Sasaki-Einstein spaces to quivers via BPS geodesics: L$^{p,q|r}$''
  JHEP {\bf 0604}, 033 (2006)
  [hep-th/0505206].
     
\bibitem{Butti:2005sw} 
  A.~Butti, D.~Forcella and A.~Zaffaroni,
  ``The Dual superconformal theory for L$^{p,q|r}$ manifolds,''
  JHEP {\bf 0509}, 018 (2005)
  [hep-th/0505220].

\bibitem{Halmagyi:2012ic} 
  N.~Halmagyi, K.~Pilch and N.~P.~Warner,
  ``On Supersymmetric Flux Solutions of M-theory,''
  arXiv:1207.4325 [hep-th].

\bibitem{Gulotta:2011si} 
  D.~R.~Gulotta, C.~P.~Herzog and S.~S.~Pufu,
  ``From Necklace Quivers to the F-theorem, Operator Counting, and T(U(N)),''
  JHEP {\bf 1112}, 077 (2011)
  [arXiv:1105.2817 [hep-th]].

\bibitem{Gulotta:2011aa} 
  D.~R.~Gulotta, C.~P.~Herzog and S.~S.~Pufu,
  ``Operator Counting and Eigenvalue Distributions for 3D Supersymmetric Gauge Theories,''
  JHEP {\bf 1111}, 149 (2011)
  [arXiv:1106.5484 [hep-th]].

\end{thebibliography}
\end{document}